\documentclass[conference]{IEEEtran}
\usepackage[utf8]{inputenc}
\usepackage{graphicx}
\usepackage{amsmath}
\usepackage{listings}
\usepackage{rotating}
\usepackage{proof}
\usepackage{color}
\usepackage{amssymb}
\usepackage{url}
\usepackage{multirow}
\usepackage{subfigure}
\usepackage{epic,eepicemu}
\usepackage{version}

\definecolor{dgreen}{rgb}{0.13,0.54,0.13}

\newtheorem{definition}{Definition}
\newtheorem{lemma}{Lemma}
\newtheorem{example}{Example}
\newtheorem{theorem}{Theorem}

\newtheorem{verification-assumption}{Verification assumption}
\newtheorem{framework-assumption}{Framework assumption}

\newtheorem{property}{Property}
\newcommand{\close}{\IEEEQEDclosed}

\newcommand{\secref}[1]{Section~\ref{#1}}

\newcommand{\fix}[2]{} 

\includeversion{LONG}
\excludeversion{SHORT}

\begin{document}

\begin{SHORT}
\title{Verifying the Interplay of Authorization Policies and Workflow in Service-Oriented Architectures}
\end{SHORT}

\begin{LONG}
\title{Verifying the Interplay of Authorization Policies and Workflow in Service-Oriented Architectures (Full version)}
\end{LONG}

\author{\IEEEauthorblockN{Michele Barletta \qquad Silvio Ranise \qquad Luca Vigan\`o}
\IEEEauthorblockA{Department of Computer Science, University of Verona, Italy \\
\{\url{michele.barletta} $\mid$ \url{silvio.ranise} $\mid$ \url{luca.vigano}\}\url{@univr.it}}
}

\maketitle

\begin{abstract}
  A widespread design approach in distributed applications based on
  the service-oriented paradigm, such as web-services, consists of
  clearly separating the enforcement of authorization policies and the
  workflow of the applications, so that the interplay between the
  policy level and the workflow level is abstracted away.  While such
  an approach is attractive because it is quite simple and permits one
  to reason about crucial properties of the policies under
  consideration, it does not provide the right level of abstraction to
  specify and reason about the way the workflow may interfere with the
  policies, and vice versa.
  \begin{LONG}
    For example, the creation of a certificate as a side effect of a
    workflow operation may enable a policy rule to fire and grant
    access to a certain resource; without executing the operation, the
    policy rule should remain inactive.  Similarly, policy queries may
    be used as guards for workflow transitions.
  \end{LONG}

  In this paper, we present a two-level formal verification framework
  to overcome these problems 
  and formally reason about the interplay of authorization policies and
  workflow in service-oriented architectures.  This allows us to
  define and investigate some verification problems for SO
  applications and give sufficient conditions for their decidability.
\end{abstract}

\section{Introduction}

A widespread design approach 
in distributed applications based on the
Service-Oriented paradigm (SO), such as web-services, consists of
clearly separating the Workflow (WF) from the Policy Management (PM).
The former orchestrates complex processing of data performed by the
various principals using a set of resources made available in the
application, while the latter aims to regulate access decisions to the
shared resources, based on policy statements made by the involved
principals. This separation of concerns is beneficial in several
respects for the design, maintenance, and verification of the
resulting applications such as reusing policies across applications.

One of the key problems in obtaining a correct design of SO
applications is to be able to foresee all the---sometimes
subtle---ways in which their WF and PM levels interact. To understand
the difficulty underlying this endevor, let us first consider the WF
level. In this respect, SO applications can be seen as distributed
systems whose transitions can be interleaved in many possible
ways. This already creates a first 
difficult problem: to understand the behaviors of an SO application
and then to establish if it meets certain properties. An additional
burden, typical to SO applications, is the presence of the PM level,
which is supposed to constrain the allowed behaviors of the
application so as to meet certain crucial security
requirements. Declarative policy languages (such as Datalog and other
languages built on top of it, like Binder~\cite{deTreville:Binder},
SecPal~\cite{SecPAL-homepage} and DKAL~\cite{DKAL-homepage}), usually
based on a (fragment of) first-order logic, are used to design the PM
level of SO applications in a more flexible, reusable, and
verification-friendly way. The high flexibility and expressiveness of
such languages may grant access to a resource to someone who, in the
intention of the policy designer, is not allowed to do so.

To further complicate the situation, there are the subtle ways in which the WF and the PM levels may interact so as to give rise to behaviors that are unintended and may breach some crucial security requirements of an SO application. As a concrete example of this point, consider a system for virtual Program Committee meetings. A policy governing access to the reviews of a paper may be the following: a reviewer assigned to a paper is required to submit his own review before being able to read those of the others.  So, in order to resolve an access request to the reviews of a paper, the system should be able not only to know the identities and the roles of the various members of the Program Committee but also to maintain and consult the information about which reviewers have already submitted their reviews. Indeed, information of the first kind must be derived from the WF level.

Given all the difficulties to obtain correct designs for SO
applications, formal methods have been advocated to help in this
task. Unfortunately, most (see, e.g.,~\cite{fisler06,ryan05}) of the
specification and verification techniques (with some notable
exceptions, e.g.,~\cite{schaad}) have concentrated on one level at a
time and abstracted away the possible interplays between the WF and the PM
level. The \emph{first contribution} of this paper is a framework
capable of formalizing both the WF and the PM level as well as their
interface so as to enable a more precise analysis of the possible
behaviors of SO applications.  In particular, we use a temporal
extension of first-order logic, similarly to what has been proposed
in~\cite{manna-pnueli-book} for the specification and verification of
reactive systems.  The motivations for this choice are
three-fold. First, workflows can be easily specified by using
first-order formulae to describe sets of states and transitions of SO
applications. Second, a simple extension of this well-known framework
allows us to easily specify policy-relevant facts and
statements. Third, we hope to adapt and reuse to the case of SO
applications the cornucopia of specification and verification
techniques developed for reactive systems.  As a first step in this
direction, the \emph{second contribution} of this paper is to define
and investigate some verification problems for SO applications and
give sufficient conditions for their decidability. In particular, we
show how executability and some security properties (which can be
expressed as invariants) can be automatically verified within the
proposed framework.

We proceed as follows. In Section~\ref{subsec:comb-fol-ltl}, we
summarize the key points of a restricted combination of first-order
and temporal logic, which provides a formal basis for our approach.
In Section~\ref{sec:framework}, we present our formal two-level
specification framework for SO applications, which we apply, in
Section~\ref{sec:verification}, on a number of interesting
verification problems for SO applications. In
Section~\ref{sec:conclusions}, we discuss related and future work, and
draw conclusions.
\begin{SHORT}
Due to lack of space, proofs are given in the accompanying technical report~\cite{BRV-TR09}, which also contains the detailed formalization of a case study that illustrates our framework at work, as well as a number of useful pragmatical observations.
\end{SHORT}
\begin{LONG}
Due to lack of space, proofs are given in the appendix, together with 
the detailed formalization of a case study that illustrates our framework at work, and with a number of useful pragmatical observations.
\end{LONG}

\section{A restricted combination of first-order logic and temporal logic}
\label{subsec:comb-fol-ltl}

As a formal basis for our approach we use a
standard~\cite{manna-pnueli-book} minimal extension of \emph{Linear Time Logic} (\emph{LTL})
with a \emph{many-sorted version of First-Order Logic with equality}
(\emph{FOL$=$}). We recall now some useful definitions and properties of
FOL$=$ where, for brevity, we do not explicitly consider sorts
although all notions can be easily adapted to the many-sorted version.
We assume the usual first-order syntactic notions of \emph{signature},
\emph{term}, \emph{literal}, \emph{formula}, \emph{quantifier-free
  formula}, \emph{substitution} and \emph{grounding} substitution,
\emph{sort} and so on, and call \emph{sentence} a formula that does
not contain free variables. Also the semantic notions of
\emph{structure}, \emph{satisfiability}, \emph{validity}, and
\emph{logical consequence} are the standard ones.  

Let $\Sigma$ be a FOL$=$ signature. 
An \emph{expression} is a term, an atom, a literal, or a formula.
A \emph{$\Sigma(\underline{x})$-expression} is an expression built out
of the symbols in $\Sigma$ where at most the variables in
the sequence $\underline{x}$ of variables may occur free, and we write
$E(\underline{x})$ to emphasize that $E$ is a
$\Sigma(\underline{x})$-expression. Similarly, for a finite sequence
$\underline{r}$ of predicate symbols in $\Sigma$, we write
$\phi(\underline{r})$ to denote a $\Sigma$-formula where at most the
predicate symbols in $\underline{r}$ may occur.  We juxtapose
sequences to denote their concatenation,
e.g.~$\underline{x}\underline{y}$, and abuse notation and write
$\emptyset$ to denote the empty sequence besides the empty set.  If
$\sigma$ is a substitution and $\underline{t}$ is a (finite) sequence
of expressions, then $\underline{t}\sigma$ is the sequence of
expressions obtained from $\underline{t}$ by applying the substitution
$\sigma$ to each element of $\underline{t}$.  

Following~\cite{manna-pnueli-book}, we use a tuple $\underline{x}$ of
variables, called \emph{WF state variables}, to represent the values
of application variables at a given instant of time, and use a FOL
formula $\varphi(\underline{x})$ to represent sets of states.  WF
state variables take values in the domain of a first-order structure,
which formalizes the data structures, the values of the control
locations, and those of the auxiliary variables of the WF of a certain
SO application.  Formally, let $\Sigma$ be a signature (containing,
e.g., the operators of certain data structures or the names of some
control locations) and $\mathcal{M}$ be a $\Sigma$-structure;
$\mathcal{M}, v\models \varphi(\underline{x})$ means that the state
formula $\varphi(\underline{x})$ is true in $\mathcal{M}$ for the
valuation $v$ mapping the variables in $\underline{x}$ to elements
of the domain of $\mathcal{M}$.  As shown in~\cite{manna-pnueli-book},
this is enough for the specification of virtually any reactive system
and hence also for the WF level.  However, the state of SO
applications should also support the PM level whose relevant part is
represented by tables where certain facts are recorded (e.g.,
``is-reviewer-of'' for the example in the introduction).  Following
the relational model of databases, we formalize tables as predicates
and we add to the WF state variables a set $\underline{p}$ of fresh
predicate symbols (i.e.\ $\underline{p}\cap \Sigma=\emptyset$), called
\emph{PM state variables}.  Any $\Sigma$-formula
$\varphi(\underline{x},\underline{p})$ is an \emph{SO state formula}. 
For a $\Sigma$-structure $\mathcal{M}=(I,D)$, a valuation $v$ mapping
the WF state variables in $\underline{x}$ to elements of the
domain $D$, and a relational valuation $b$ mapping the PM state
variables in $\underline{p}$ (such that $\underline{p}\cap
\Sigma=\emptyset$) to the powerset of $D$, we write 
\begin{displaymath}
\mathcal{M}, v, b\models \varphi(\underline{x},\underline{p})
\end{displaymath}
to denote that $\mathcal{M}_{b},
v\models \varphi(\underline{x},\underline{p})$ where
$\mathcal{M}_{b}=(I',D')$ is the $(\Sigma\cup
\underline{p})$-structure obtained from $\mathcal{M}$ by taking
$D'=D$, $I'|_{\Sigma} = I$, and $I'(p)=b(p)$ for each $p\in
\underline{p}$.  The tuple $(\mathcal{M},v,b)$ (or simply $v,b$, when
$\mathcal{M}$ is clear from context) is an \emph{SO state}.

Let $\Sigma$ be a signature and $\mathcal{M}$ be a $\Sigma$-structure.
We formalize an SO application by a tuple
$(\underline{x},\underline{p}, \iota, \mathit{Tr})$, called an
\emph{SO transition system}, where $\underline{x}$ are the WF state
variables, $\underline{p}$ are the PM state variables, $\iota$ is a
$\Sigma(\underline{x},\underline{p})$-formula, and $\mathit{Tr}$ is a
finite set of
$\Sigma(\underline{x},\underline{p},\underline{x}',\underline{p}')$-formulae,
called \emph{transitions}, which relate a set of SO states (identified
by the ``values'' of $\underline{x},\underline{p}$) to that of a set
of SO ``next'' states (identified by the ``values'' of
$\underline{x}',\underline{p}'$).\footnote{We ignore fairness
  assumptions as, for simplicity in this paper, we are only concerned
  with security properties that can be encoded as a sub-class of
  safety properties.}  If $\underline{p}=\emptyset$ and
$\underline{x}\neq \emptyset$, then our notion of SO transition system
reduces to that of transition system in~\cite{manna-pnueli-book}; in
the rest of this paper, we assume that $\underline{p}\neq \emptyset$.

A \emph{run} of an SO transition system $(\underline{x},\underline{p},
\iota, \mathit{Tr})$ is an infinite sequence of SO states $v_0,b_0,
..., v_i,b_i, ...$ such that $\mathcal{M}, v_0,b_0 \models
\iota(\underline{x},\underline{p})$ and for every $i\geq 0$, there
exists a transition
$\tau(\underline{x},\underline{p},\underline{x}',\underline{p}')\in
\mathit{Tr}$ such that $\mathcal{M}, v_i,b_i,v_{i+1},b_{i+1} \models
\tau(\underline{x},\underline{p},\underline{x}',\underline{p}')$ where
$v_i,b_i$ (respectively, $v_{i+1},b_{i+1}$) map state variables and predicates in
$\underline{x},\underline{p}$ (respectively, $\underline{x}',\underline{p}'$).

To specify properties of SO transition systems, we use an extension of
Linear Time Logic. Formally, let
$\underline{x},\underline{p}$ be SO state variables and $\Sigma$ be a
signature; the set $\mathit{LTL}(\Sigma,\underline{x},\underline{p})$
of \emph{LTL (state-based) formulae for $\Sigma$ and
  $\underline{x},\underline{p}$} is inductively defined as
follows:
state formulae are in $\mathit{LTL}(\Sigma,\underline{x},\underline{p})$ and if $\varphi$ is
a state formula then $\Box \varphi$ is in
$\mathit{LTL}(\Sigma,\underline{x},\underline{p})$.\footnote{The
  minimalist temporal logic defined here suffices for the purposes of
  specifying the sub-class of invariant properties that are relevant
  for this paper.  However, the proposed framework may be easily
  extended to support other temporal operators such as ``sometimes in
  the future,'' ``next'', or ``until.''} Note that we prohibit
alternation of FOL quantifiers and temporal operators: this makes the
logic less expressive but it helps to derive decidability results for
the satisfiability problem, which is a necessary condition to develop
(semi-)automatic verification methods for SO applications.

Let $\mathcal{M}$ be a $\Sigma$-structure.  A \emph{model} of an
$\mathit{LTL}(\Sigma,\underline{x},\underline{p})$-formula is an
infinite sequence $v_0,b_0, ..., v_i,b_i, ...$ of SO states such that
each $v_i,b_i$ map all the SO state variables in
$\underline{x},\underline{p}$, for $i\geq 0$.
We then say that an
$\mathit{LTL}(\Sigma,\underline{x},\underline{p})$-formula
$\varphi(\underline{x},\underline{p})$ is \emph{true in a model
$v_0,b_0, ..., v_i,b_i, ...$}, and write
\begin{displaymath}
\mathcal{M}, v_0,b_0,..., v_i,b_i, ... \models \varphi(\underline{x},\underline{p})\,,
\end{displaymath}
iff
\begin{itemize}
\item $\mathcal{M}, v_0, b_0 \models \varphi$ whenever
  $\varphi(\underline{x},\underline{p})$ is a state formula;
\item $\mathcal{M}, v_0,b_0, ..., v_i,b_i, ...  \models \Box
  \varphi(\underline{x},\underline{p})$ iff $\mathcal{M}, v_k, b_k
  \models \varphi(\underline{x},\underline{p})$, for every $k\geq 0$.
\end{itemize}

Let $\mathcal{S}=(\underline{x},\underline{p}, \iota, \mathit{Tr})$ be
an SO transition system, $\mathcal{M}$ be a $\Sigma$-structure, and
$\varphi(\underline{x},\underline{p})$ be an
$\mathit{LTL}(\Sigma,\underline{x},\underline{p})$-formula.  Then,
$\mathcal{S}\models \varphi(\underline{x},\underline{p})$ iff
$\mathcal{M},v_0, b_0, ..., v_i,b_i, ... \models
\varphi(\underline{x},\underline{p})$ for every run $v_0, b_0, ...,
v_i,b_i, ... $ of $\mathcal{S}$.
A state formula $\psi(\underline{x},\underline{p})$ is an
\emph{invariant for the SO transition system $\mathcal{S}$} if
$\mathcal{S}\models \Box \psi(\underline{x},\underline{p})$.

\section{A formal two-level specification framework  for SO applications}\label{sec:framework}
\begin{figure}[tb]
\centering
\vspace*{-0.2cm}
\includegraphics[scale=.6]{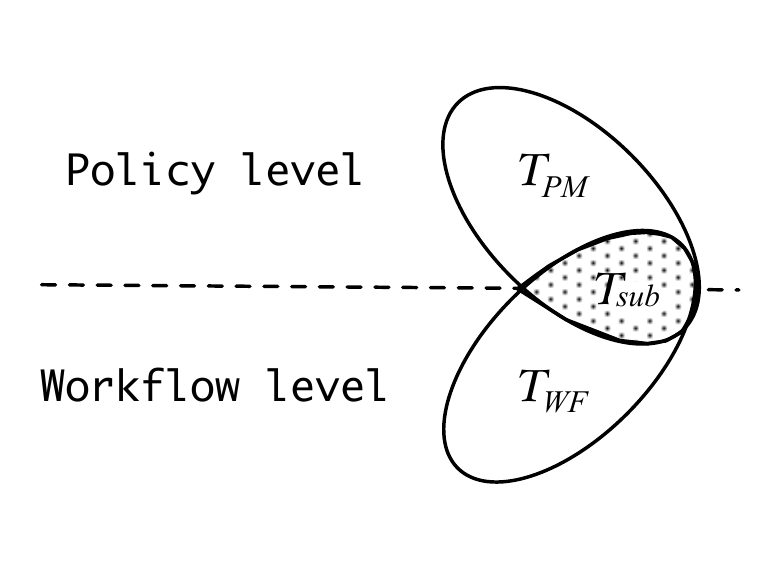}
\vspace*{-0.5cm}
\caption{\label{fig:ths-rels}FOL formalization of the WF and PM levels
  of SO applications}
\end{figure}

Recall that one of the main goals of this paper is to
provide a natural specification framework for SO applications whose
architecture is organized in two levels.  Indeed, it is possible to
model a large class of SO applications by using the notion of SO
transition system introduced in the previous section.  However, a good
framework should provide an adequate support to restrict the design
space for a two-level SO application and allow the designer to easily
specify the WF level, the PM level, and their interface in isolation
according to a divide-and-conquer strategy.  In our framework, this
consists of identifying suitable first-order structures formalizing
both the data structures at the WF level and the tables at the PM
level.  Unfortunately, working with first-order structures for
specification is quite difficult since there is no obvious way to
mechanically represent and reason about them.  Fortunately,
first-order theories are sets of FOL$=$ sentences that can be used as
reasonably precise approximations of first-order structures and for
which there is automated reasoning support.  Hence, we decided to use
theories to describe the WF, the PM levels, and their interface, as illustrated in 
Fig.~\ref{fig:ths-rels}: if $T_\mathit{WF}$ and $T_\mathit{PM}$ are the theories 
formalizing the WF and the PM levels, respectively, then their intersection $T_\mathit{sub} = T_\mathit{WF} \cap
T_\mathit{PM}$, called the \emph{substrate theory}, formalizes their interface.

Intuitively, the theory $T_\mathit{sub}$ ensures that the WF and PM levels
``agree'' on certain notions. For example, $T_\mathit{sub}$ univocally
identifies the principals involved in the SO application and possibly
(an abstraction of) the {structure} of the resources that the SO
application can access or make available.  As we will see, the use of
theories allows us to easily import declarative policy specifications
expressed in logical languages based on (extensions of) Datalog in our
framework.
\label{sec:modelling}

A similar approach can also be used to restrict the formulae
characterizing transitions.  Intuitively, the transitions that can be
specified by formulae in the identified class are such that the
updates on the values of both the WF and the PM state variables are
functional (if we regard relations as first-order objects).  Since the
identities of the principals involved in the SO application being
specified play a crucial role in enabling or disabling the possibility
to execute a certain transition, the functional updates will depend
not only on the values of the actual SO state but also on the
existence of certain principals.

Before being able to formalize these intuitions, we recall the concept
of FOL$=$ theory and some standard related notions.

\subsection{First-order theories for SO applications}
\label{subsec:fol-theories-for-SOAs}
A \emph{$\Sigma$-theory} $T$ is a set of first-order
$\Sigma$-sentences, called the \emph{axioms} of $T$.  A
$\Sigma$-structure $\mathcal{M}$ is a \emph{model} of the
$\Sigma$-theory $T$ iff all the sentences in $T$ are true in
$\mathcal{M}$.  A $\Sigma$-theory $T$ is \emph{consistent} if it
admits at least one model.
The $T$-\emph{satisfiability problem} for a quantifier-free
$\Sigma$-formula $\varphi(\underline{x},\underline{p})$ (such that
$\underline{p}\cap \Sigma=\emptyset$) consists of checking whether
there exists a model $\mathcal{M}$ of $T$ and mappings $v$ and $b$
such that $\mathcal{M},v,b \models
\varphi(\underline{x},\underline{p})$.  By transformation in
disjunctive normal form (i.e.~as disjunctions of conjunctions of
literals), the $T$-satisfiability problem for quantifier-free formulae
can be reduced to the $T$-satisfiability problem for (quantifier-free)
conjunctions of literals.

For specifying SO applications, we usually need to introduce a finite
set of unique identifiers to name the various principals.  Formally,
this can be done by using a theory of the following kind.  An
\emph{enumerated data-type theory $\mathit{EDT}(C,S)$} is axiomatized
by the sentences
\begin{displaymath}
  \bigwedge_{c_i,c_j\in C, i\neq j} c_i\neq c_j \qquad \text{and} \qquad
  \forall x:S.\, \bigvee_{c\in C} x = c \,,
\end{displaymath}
for $S$ a sort symbol in the given signature (omitted for simplicity)
and $C=\{c_1, ..., c_n\}$, $c_i$ of sort $\mathit{S}$ for $i=1,
\ldots, n$ and $n\geq 1$.  It is easy to see that the
$\mathit{EDT}(C,S)$-satisfiability problem is decidable.  Enumerated
data-type theories will be sub-theories of the theory formalizing the
interface between the WF and PM levels.

For the WF level, we can reuse all the theories available in the
literature formalizing data structures and the decision procedures for
their satisfiability problem (see, e.g.,~\cite{sebastiani} for an
overview).  Enumerated data-type theories are also useful for the WF
level as they can formalize the (finitely many) control locations of
an application.  We now give a concrete example of a theory capable of
formalizing a simple message passing network that is relevant for SO
applications 
\begin{SHORT}
(see~\cite{BRV-TR09} for a more detailed case study).
\end{SHORT} 
\begin{LONG}
(see Appendix~\ref{subsec:running-ex} for a more detailed case study).
\end{LONG}
\begin{example}[Message passing]
  \label{ex:msg-pass}
  A net can be seen abstractly as a set of messages: sending a message
  amounts to adding the message to the set while receiving a message
  consists of checking if it is a member of the net; hence, messages
  are never deleted, only added to the set representing the net.  This
  view is simple but still allows one to model interesting facts such
  as the reception of messages in any order (since a set does not
  require an ordering on its elements) or duplication of messages (as
  a message is never removed from the net).

  To model the simple fragment of set theory necessary to formalize
  this idea in FOL, we use a theory $\mathit{MsgPass}[\mathit{Msg}]$,
  parametrized over the sort $\mathit{Msg}$ of messages which contains
  $\mathit{SetOfMsg}$ as the sort for sets of messages, the constant
  $\mathsf{mty}$ of sort $\mathit{SetOfMsg}$ denoting the empty set,
  the binary function symbol $\mathsf{ins}$ of sort $\mathit{Msg}
  \times \mathit{SetOfMsg} \rightarrow \mathit{SetOfMsg}$ denoting the
  operation of adding a message to a set of messages, and the binary
  predicate symbol $\mathsf{mem}$ of sort $\mathit{Msg} \times
  \mathit{SetOfMsg}$ for checking if a message is in a set of
  messages. The axioms of $\mathit{MsgPass}[\mathit{Msg}]$ are the
  following three sentences:
  \begin{eqnarray*}
    \forall E.\, \neg \mathsf{mem}(E, \mathsf{mty}) \quad\quad
    \forall E.\, \mathsf{mem}(E, \mathsf{ins}(E,S)) \\
    \forall E,E'.\, E\neq E' \rightarrow 
    (\mathsf{mem}(E, \mathsf{ins}(E',S)) \leftrightarrow \mathsf{mem}(E,S))
  \end{eqnarray*}
  where $E,E'$ are variables of sort $\mathit{Msg}$ and $S$ is a
  variable of sort $\mathit{SetOfMsg}$.  It is easy to describe the
  states of a variable \texttt{net}: just introduce a logical variable
  $\mathit{net}$ and use suitable formulae from the theory
  $\mathit{MsgPass}[\mathit{Msg}]$.  For example,
  the formula $\exists m_1,m_2 : \mathit{Msg}, \mathit{net}' :
  \mathit{SetOfMsg}.\, m_1\neq m_2 \wedge \mathit{net} =
  \mathsf{ins}(m_1,\mathsf{ins}(m_2,\mathit{net}'))$ constrains
  \texttt{net} to contain at least two messages (plus possibly
  others).  Note that the only free variable in the formulae
  describing sets of states is $\mathit{net}$.  The
  $\mathit{MsgPass}[\mathit{Msg}]$-satisfiability problem is
  decidable~\cite{ic03}.  \hfill \close
\end{example}

For the PM level, we recall the class of Bernays-Sch\"onfikel-Ramsey
(BSR) sentences~\cite{Boerger97}, which has been used, among other applications, to
model relational databases.  A \emph{BSR-theory} is a
set of sentences of the form
\begin{displaymath}
  \exists \underline{x}~\forall \underline{y}.\, 
    \varphi(\underline{x},\underline{y}) ,
\end{displaymath}
where $\underline{x}$ and $\underline{y}$ are tuples of variables and
$\varphi$ is a quantifier-free formula containing only predicate and
constant symbols (or, equivalently, no function symbols of arity
greater than or equal to $1$).  The decidability of the satisfiability
problem for any BSR-theory is a well-known decidability
result~\cite{Boerger97}.

The following sub-class of BSR-theories can be used to specify
policies as shown in, e.g.,~\cite{SPKI-SDSI-in-FOL}.  A
\emph{Datalog-theory} is a BSR-theory whose sentences are of the form
\begin{equation}\label{eq:horn-cl}
    \forall \underline{x},\underline{y}.\,
     q_1(\underline{x}, \underline{y}) \wedge \cdots \wedge
     q_n(\underline{x}, \underline{y}) 
      \ \rightarrow \
      p(\underline{x})
\end{equation}
where $p,q_i$, for $i=1,...,n$, are predicate symbols,
and $\underline{x},\underline{y}$ are disjoint tuples of variables such
that the length of $\underline{x}$ is equal to the arity of $p$.
Usually, sentences of the form (\ref{eq:horn-cl}) are written as 
\begin{displaymath}
  \forall \underline{x},\underline{y}.\,
  p(\underline{x})
  \ \leftarrow \
  q_1(\underline{x}, \underline{y}) \wedge \cdots \wedge
  q_n(\underline{x}, \underline{y}) \,,
\end{displaymath}
where $\leftarrow$ can be read as the reverse of the implication
connective (sometimes also the universal quantifiers will be dropped).
Formulae written in this way are called \emph{rules} in the
literature, while their hypotheses and $p(\underline{x})$ are called
the \emph{body} and the \emph{head} of the rule, respectively.

We conclude by recalling some notions that are relevant for the
combination of theories that 
provide us with the formal tools to separately specify the WF and PM
levels of an SO application and then modularly combine them.  Let
$T_1$ and $T_2$ be two theories; we say that they \emph{share} the
theory $T_0=T_1\cap T_2$ if $T_0\neq \emptyset$ and their
\emph{combination} $T_1\cup T_2$ is \emph{non-disjoint}.  Otherwise
(i.e.\ when $T_0=\emptyset$), we say that the {combination} $T_1\cup
T_2$ is \emph{disjoint}.  For verification, it is important to combine
decision procedures for each theory $T_1$ and $T_2$ so as to obtain a
decision procedure for their combination.  This is crucial to 
derive decidability results for verification problems of SO
applications as we will reduce them to satisfiability problems in the
combination of the theories formalizing the WF and the PM levels.  A
class of theories that will be relevant in this task (see
Lemma~\ref{th:dec-SOA-sat-univ} below) is the following.  A theory $T$
is \emph{stably infinite} if a $T$-satisfiable quantifier-free formula
is satisfiable in a model of $T$ whose domain has infinite
cardinality.  Examples of stably infinite theories are
$\mathit{MsgPass}[\mathit{Msg}]$ of Example~\ref{ex:msg-pass}, any BSR
theory (see, e.g.,~\cite{tinelli-zarba}), and many theories
formalizing data structures, such as arrays or sets.  Enumerated
data-type theories are not stably infinite as they admit only models
whose domains have finite cardinality.

\subsection{Two-level SO transition systems}
We are now ready to define an instance of the framework
of~\secref{subsec:comb-fol-ltl} to formally specify SO
applications designed according to the two-level architectures
considered in this paper. This framework relies on the following assumptions.
\begin{framework-assumption}\label{framework-one}
  As depicted in Fig.~\ref{fig:ths-rels}, we assume that the WF and
  PM levels are formalized by a $\Sigma_\mathit{WF}$-theory
  $T_\mathit{WF}$ and a $(\Sigma_\mathit{PM}\cup
  \underline{p})$-theory $T_\mathit{PM}$, which share a
  $\Sigma_\mathit{sub}$-theory $T_\mathit{sub}$, called the
  \emph{substrate}. Formally, $\Sigma_\mathit{sub}\subseteq
  \Sigma_\mathit{WF}$ and $\Sigma_\mathit{sub}\subseteq
  \Sigma_\mathit{PM}$, and $T_\mathit{sub}\subseteq T_\mathit{WF}$ and
  $T_\mathit{sub}\subseteq T_\mathit{PM}$. \hfill \close
\end{framework-assumption}

The shared theory $T_\mathit{sub}$ plays the role of \emph{interface} between the
two levels.  A minimal requirement on the interface is to provide some
knowledge about the identities of the principals involved in the SO
application. This is formalized as follows.
\begin{framework-assumption}\label{framework-two}
  $\Sigma_\mathit{sub}$ contains the sort symbol $\mathit{Id}$. \hfill \close
\end{framework-assumption}

This last assumption is crucial for many aspects of SO applications
related to PM, such as integrity (of messages or certificates),
authenticity (of certificates), and proof-of-compliance (of
credentials).

Using the notion of combination of theories introduced at the end
of~\secref{subsec:fol-theories-for-SOAs}, we are now able to define
the concept of background theory for an SO application that is obtained by
modularly combining the theories formalizing the WF and the PM levels.
Let $T_\mathit{sub}, T_\mathit{WF}$, and $T_\mathit{PM}$ be consistent
theories satisfying Framework assumptions~\ref{framework-one}
and~\ref{framework-two}.  The \emph{background
  $\Sigma_\mathit{SOA}$-theory $T_\mathit{SOA}$} is the union of the
theories $T_\mathit{WF}$ and $T_\mathit{PM}$, i.e.\
$\Sigma_\mathit{SOA}:= \Sigma_\mathit{WF} \cup \Sigma_\mathit{PM}$
and $T_\mathit{SOA}:=T_\mathit{WF}\cup T_\mathit{PM}$.  Note that, by
Robinson consistency theorem (see, e.g.,~\cite{chang-kiesler}),
$T_\mathit{SOA}$ is consistent since both $T_\mathit{WF}$ and
$T_\mathit{PM}$ are assumed to be so.  We will sometimes refer to
$T_\mathit{WF}$ as the \emph{WF background theory} and to
$T_\mathit{PM}$ as the \emph{PM background theory}.

We introduce a particular class of SO transition systems (defined
in~\secref{subsec:comb-fol-ltl}) by using background theories obtained
by combining theories for the WF and PM levels satisfying the two
framework assumptions above.  A technical problem in doing this is the
following.  SO transition systems (in particular their states and
runs) are defined with respect to a certain first-order structure.  Instead,
we want to use theories that, in general, identify classes of
first-order structures and not just one particular structure.
However, since the verification problems for SO applications
considered below will be reduced to satisfiability problems, the
following notion tells us that---under suitable conditions---we can
use theories in place of structures.  A $\Sigma$-theory $T$ is
\emph{adequate} for a $\Sigma$-structure $\mathcal{M}$ if 
$\mathcal{M}, v\models \varphi(\underline{x})$, for some valuation $v$
mapping the variables in $\underline{x}$ to elements of the domain of
$\mathcal{M}$, is equivalent to the $T$-satisfiability of
$\varphi(\underline{x})$, for any quantifier-free formula
$\varphi(\underline{x})$.  For example, it is possible to see that
enumerated data-type theories are adequate for any of their models (as
they are all isomorphic) or that the theory
$\mathit{MsgPass}[\mathit{Msg}]$ is adequate to the structure
containing finite sets of messages.  

As notation, let us write $\forall \underline{z}.\, \underline{p}(\underline{z}) \leftrightarrow \underline{\iota_{PM}}(\underline{i}, \underline{p}, \underline{z})$ (respectively, $\forall \underline{z}.\, \underline{p}'(\underline{z}) \leftrightarrow \underline{\varphi}(\underline{i}, \underline{p}, \underline{z})$) to abbreviate the finite conjunction, for $j=1,...,n$, of formulae of the form $\forall \underline{z}_j.\, p_j(\underline{z}_j) \leftrightarrow
 \iota_j(\underline{i},\underline{p},\underline{z}_j)$ (respectively, $\forall \underline{z}_j.\, p_j(\underline{z}_j)
 \leftrightarrow \varphi_j(\underline{i},\underline{p},\underline{z}_j)$) when $\underline{p} = p_1, \ldots, p_n$, $\underline{\iota_{PM}} = \iota_1, \ldots, \iota_n$ (respectively, $\underline{\varphi} = \varphi_1, \ldots, \varphi_n$), $\underline{z}_j\subseteq \underline{z}$, and the length of $\underline{z}_j$ is equal to the arity of $p_j$.
  
\begin{definition}[Two-level SO transition system]
  Let $T_\mathit{sub}$, $T_\mathit{WF}$, and $T_\mathit{PM}$ be consistent
  theories satisfying Framework assumptions~\ref{framework-one} and~\ref{framework-two} 
  and $\mathcal{M}$ be
  a $\Sigma_\mathit{SOA}$-structure.  A \emph{two-level SO
    transition system (with background theory $T_\mathit{SOA}$ 
    adequate for $\mathcal{M}$)} is an SO transition system
  $(\underline{x},\underline{p},\iota,\mathit{Tr})$ such that (a)
  $\underline{p}\cap \Sigma_\mathit{SOA}=\emptyset$; (b) $\iota$ is a state
  $\Sigma_\mathit{SOA}$-formula of the form:
  \begin{eqnarray}
    \label{eq:initial-formula}
    \forall \underline{i}.\,( 
    \iota_{WF}(\underline{i},\underline{x}) \, \wedge \,
    \forall \underline{z}.\,
    \underline{p}(\underline{z}) \leftrightarrow 
    \underline{\iota_{PM}}(\underline{i},\underline{p},\underline{z})) \,,
  \end{eqnarray}
  where $\underline{i}$ is a finite sequence of variables of sort
  $\mathit{Id}$, $\iota_{WF}$ is a quantifier-free
  $\Sigma_{WF}(\underline{i},\underline{x})$-formula, and $\iota_{PM}$
  is a quantifier-free
  $\Sigma_{PM}(\underline{z},\underline{i},\underline{p})$-formula; and (c)
  $\mathit{Tr}$ is a finite state of transition formula of the form
  \begin{eqnarray}
    \label{eq:trans-form}
    \exists \underline{i},\underline{d}.\, (
    G(\underline{i}, \underline{d}) 
    \wedge
    \underline{x}' = \underline{f}(\underline{x}, \underline{i}, \underline{d}) 
    \wedge
    \forall \underline{z}.\, \underline{p}'(\underline{z}) \leftrightarrow
    \underline{\varphi}(\underline{i}, \underline{p}, \underline{z}) )\,,
  \end{eqnarray}
  called \emph{guarded assignment transition}, where $\underline{i}$
  is a tuple of variables of sort $\mathit{Id}$; $\underline{d},
  \underline{z}$ are sets of variables of a sort dependent on the WF
  and PM levels of the application; $G$ is a quantifier-free formula,
  called the \emph{guard} of the transition; $\underline{f}$ is a
  tuple of $\Sigma_\mathit{WF}(\underline{x}, \underline{i})$-terms,
  called the \emph{WF updates} of the transition, whose sorts are
  pairwise equal to those of the state variables in $\underline{x}$;
  and $\underline{\varphi}$ is a tuple of quantifier-free
  $\Sigma_\mathit{PM}(\underline{i}, \underline{p},
  \underline{z})$-formulae, called the \emph{PM updates} of the transition. 
  \hfill\close
\end{definition}

If $\underline{x}=\emptyset$ (recall that we have assumed that
$\underline{p}\neq \emptyset$ for SO transition systems,
cf.~\secref{subsec:comb-fol-ltl}), then we say that the SO application
is \emph{(purely) relational}.  Intuitively, the form
(\ref{eq:initial-formula}) for the initial state formula is inspired by
the observation that usually the principals at the beginning of the
computation have some common (or no) knowledge about the facts that
are relevant to the PM level.  Note that $\underline{f}$ and
$\underline{\varphi}$ may not contain the state variables in
$\underline{x}'$ and the state predicates in $\underline{p}'$, i.e.\
updates are not recursive.  

\begin{SHORT}
Below, for simplicity, we will no more mention the
$\Sigma_\mathit{SOA}$-structure $\mathcal{M}$ and implicitly assume
that $T_\mathit{SOA}$ is adequate for $\mathcal{M}$.  To help
intuition, we illustrate the notion of two-level SO transition system
by means of a simple example (extracted from a larger one
in~\cite{BRV-TR09}).
\end{SHORT}
\begin{LONG}
Below, for simplicity, we will no more mention the
$\Sigma_\mathit{SOA}$-structure $\mathcal{M}$ and implicitly assume
that $T_\mathit{SOA}$ is adequate for $\mathcal{M}$.  To help
intuition, we illustrate the notion of two-level SO transition system
by means of a simple example (extracted from the case study in the appendix).
\end{LONG}
\begin{example}
  \label{ex:SO-appl}
  Consider a situation where the clerks of an office may send
  messages over a network.  The messages may contain, among many other
  things, certificates about their identities, roles, or capability to
  access certain resources in the organization they belong to.
  Certificates about the identities and roles are issued by a trusted
  certification authority while those about the access to a certain
  resource are issued by heads (who are clerks with this special
  right).  In order to comply with the policies of accessing
  resources, each clerk maintains a table about his/her identity,
  role, and access capability as well as about other clerks.  We
  describe a two-level SO transition system to formalize this
  situation.  

  First of all, we specify the WF background theory:
  \begin{eqnarray*}
    T_\mathit{sub} & := & \mathit{EDT}(\{ \mathtt{Ed}, \mathtt{Helen}, \mathtt{RegOffCA}, \mathtt{Res}\}, \mathit{Id}) \cup \\
           &    & \mathit{EDT}(\{ \mathtt{employee}, \mathtt{head}\}, \mathit{Role})  \\
    T_\mathit{WF}  & := & T_\mathit{sub} \cup  \mathit{MsgPass}[\mathit{Msg}] \cup \mathit{Msg} \cup \mathit{Cert} 
  \end{eqnarray*}
  where $\mathtt{Ed}$ and $\mathtt{Helen}$ are two clerks,
  $\mathtt{RegOffCA}$ is the trusted certification authority, $\mathtt{Res}$
  is a shared resource (e.g., a repository), $\mathtt{employee}$ and
  $\mathtt{head}$ are the possible roles of clerks, and
  $\mathit{MsgPass}[\mathit{Msg}]$ is the theory for message passing
  introduced in Example~\ref{ex:msg-pass}.  In particular,
  $\mathit{Msg}$ is a theory to describe the structure of messages as
  follows: a message contains a field identifying the sender, a field
  identifying the receiver, and a field carrying their contents.
  Formally, this is done by introducing two new sort symbols
  $\mathit{Body}$ and the ternary function $\mathsf{msg}$ of sort
  $\mathit{Id} \times \mathit{Body}\times \mathit{Id} \rightarrow
  \mathit{Msg}$.  Finally, $\mathit{Cert}$ is a theory to provide
  functionalities to analyze the body of messages and extract some
  relevant information: the predicate $\mathsf{cert\_of\_role}$ of
  sort $\mathit{Body}\times \mathit{Id} \times \mathit{Role}$ is
  capable of recognizing that the body of a message contains a
  certificate that its second argument is the identifier of a clerk
  whose role is that of its third argument.  For example, if
  $\mathsf{cert\_Ed\_empl}$ is a constant of sort $\mathit{Body}$
  representing the certificate that the employee $\mathtt{Ed}$ has the
  role $\mathtt{employee}$, then the message sent by $\mathtt{RegOffCA}$ to
  $\mathtt{Helen}$ containing the certificate
  $\mathsf{cert\_Ed\_empl}$ is encoded by the following term:
  $\mathsf{msg}(\mathtt{RegOffCA}, \mathsf{cert\_Ed\_empl}, \mathtt{Helen})$
  and we will also have that, e.g., 
  $\mathsf{cert\_of\_role}(\mathsf{cert\_Ed\_empl}, \mathtt{Ed},
  \mathtt{employee})$ holds.

  \begin{figure*}[t]
  \begin{eqnarray*}
    \exists i_1,i_2,c.\, 
     \left(
       \begin{array}{l}
         \mathsf{mem}(\mathsf{msg}(\mathtt{RegOffCA}, c, i_1), \mathit{net})
         ~\wedge~
         \mathsf{cert\_of\_role}(c,i_2,\mathtt{employee})  ~~\wedge~
         \mathit{net}' = \mathit{net}  ~~\wedge \\
         \forall z_1,z_2,d.\, 
           \mathsf{hasrole}'(z_1,z_2,d) \leftrightarrow 
           \left(
             \begin{array}{l}
               \mathit{if}~ (z_1=i_1 \wedge z_2=i_2 \wedge d=\mathtt{employee}) \
               \mathit{then ~ true} \
               \mathit{else} ~ \mathsf{hasrole}(z_1,z_2,d) 
             \end{array} 
           \right)
         \end{array}
       \right)
  \end{eqnarray*}
  \centerline{where $i_1,i_2,z_1,z_2$ are variables of sort
    $\mathit{Id}$, $c$ is a variable of sort $\mathit{Body}$, and $d$
    is a variable of sort $\mathit{Role}$}
  \caption{\label{fig:transition-example3}A formalization of the interplay between WF and PM levels by a guarded assignment transition (cf.\ Example~\ref{ex:SO-appl})}
  \end{figure*}

  The state of the two-level SO transition system specifying the situation
  above should contain a WF state variable $\mathit{net}$ of sort
  $\mathit{SetOfMsg}$ (containing the set of messages exchanged during
  a run of the transition system) and a PM state variable
  $\mathsf{hasrole}$ of arity $\mathit{Id} \times \mathit{Id}\times
  \mathit{Role}$ (storing the join of the tables of each clerk about
  their roles).  The initial state of the system
  should specify that no message has been exchanged over the net and
  that no role is known to the various clerks.  This can be formalized
  by a state formula as follows:
  \begin{eqnarray*}
    \mathit{net} = \mathsf{mty} & \wedge &
    \forall z_1, z_2, r.\mathsf{hasrole}(z_1, z_2, r)
    ~\leftrightarrow~ \mathit{false} \,,
  \end{eqnarray*}
  which is a formula of the form (\ref{eq:initial-formula}) by taking
  $\underline{i}=\emptyset$, $\underline{z}=\{ z_1, z_2, r \}$,
  $\mathit{net}\in \underline{x}$, and $\mathsf{hasrole}\in
  \underline{p}$.

  As an example of interplay between the WF and PM levels,
  Fig.~\ref{ex:SO-appl} shows the guarded assignment transition of the
  form (\ref{eq:trans-form}) that formalizes what happens when a
  message containing a certificate about the role of an employee (say,
  $\mathtt{Ed}$) is sent to another employee (say, $\mathtt{Helen}$)
  by the certification authority ($\mathtt{RegOffCA}$).  Note that the
  content of the state variable $\mathit{net}$ is left unchanged by
  the transition, whose only effect is to update the access table
  (represented by the predicate $\mathsf{hasrole}$) with the entry
  corresponding to the content of the received (role) certificate.
  For example, upon reception of the message containing the
  certificate $\mathsf{cert\_Ed\_empl}$, the following fact
  $\mathsf{hasrole}'(\mathtt{Helen}, \mathtt{Ed}, \mathtt{employee})$
  must hold in the next state, while for all the other triples,
  $\mathsf{hasrole}'$ has the same Boolean value of
  $\mathsf{hasrole}$.

  So far, we have specified the WF level of the SO application.  As
  anticipated above, we can define $T_\mathit{PM}$ to contain $T_\mathit{sub}$ and a
  finite set of Datalog rules that declaratively formalize the access
  policy statements of the SO application.  Since this way of
  formalizing policies has been well studied in the literature (see,
  e.g.,~\cite{deTreville:Binder}), as a simple example, we only give
  the following Datalog rule
  \begin{eqnarray*}
    \forall i.\,
    \mathsf{can\_access}(i,\mathtt{Res}) 
     & \leftarrow &
     \mathsf{hasrole}(\mathtt{Res},i,\mathtt{head}) ,
  \end{eqnarray*}
  where the variable $i$ is of sort $\mathit{Id}$ and
  $\mathsf{can\_access} \in \Sigma_\mathit{PM}$.  It says that the clerk $i$
  can access the shared resource $\mathtt{Res}$ if the latter knows
  (by retrieving the right entry in the table represented by
  $\mathsf{hasrole}$) that $i$ has the role of $\mathtt{head}$.
  \hfill \close
\end{example}

\begin{SHORT}
  In~\cite{BRV-TR09}, we discuss in detail a generalization of the above
  example inspired by an industrial application.
\end{SHORT}
\begin{LONG}
  Appendix~\ref{sec:pragmatics-modeling} contains a generalization of
  the example above inspired by an industrial application.
\end{LONG}

\section{Some verification problems for SO applications}
\label{sec:verification}
Let $\mathcal{A}=(\underline{x},\underline{p},\iota,Tr)$ be a
two-level SO transition system with background theory $T_\mathit{SOA}$
for an SO application; for brevity, we will sometimes refer to
$\mathcal{A}$ simply as SO application. We define and investigate some
verification problems for SO applications and give sufficient
conditions for their decidability.
\begin{SHORT}
In~\cite{BRV-TR09}, we also
discuss pragmatical aspects of how to implement the decision procedures.
\end{SHORT}
\begin{LONG}
In appendices~\ref{sec:pragmatics-executability} and~\ref{sec:pragmatics-invariant}, we then 
discuss pragmatical aspects of how to implement the decision procedures.
\end{LONG}

\subsection{Executability of SO applications}
\label{sec:exec-so-apps}

Symbolic execution is a form of execution where many possible
behaviors of a system are considered simultaneously. This is achieved
by using symbolic variables to represent many possible states and
executions. For each possible valuation of these variables, there is a
concrete system state that is being indirectly simulated.  This
technique is particularly useful for the design of SO applications when
usually several scenarios are identified as typical execution paths
that the application should support.  Given the high degree of
non-determinism and the subtle interplay between the WF and
the PM levels, it is often far from being obvious that the SO
application just designed allows one or many of the chosen scenarios.  A
valuable contribution of the proposed framework is that symbolic
execution of SO applications can be done by using existing techniques
for automated deduction.  

In any scenario, there is only a known and finite number of principals. So, for the verification of the executability of two-level SO transition systems, 
we can assume that:
\begin{verification-assumption}
$T_\mathit{sub} \supseteq EDT(\underline{c}, \mathit{Id})$. \hfill \close
\end{verification-assumption}
Since there are only finitely many principals, universal quantifiers in initial state formulae do not add to expressiveness as $\forall \underline{i}.\, \iota(\underline{i},\underline{x},\underline{p})$ is logically $T_\mathit{SOA}$-equivalent to a quantifier-free formula of the form $\bigwedge_{\sigma} \iota(\underline{i}\sigma,\underline{x},\underline{p})$,
where $\sigma$ ranges over all possible grounding substitutions mapping the variables in $\underline{i}$ to the constants in $\underline{c}$.  Thus, we can further assume that:
\begin{verification-assumption}
Initial state formulae as well as any other state formula used to describe a state of a scenario are quantifier-free. \hfill \close
\end{verification-assumption}

The key notion for symbolic execution in our framework is the following. Let $\varphi(\underline{p},\underline{x})$ and $\psi(\underline{p},\underline{x})$ two quantifier-free state $\Sigma_\mathit{SOA}$-formulae; and let $\tau(\underline{p},\underline{x},\underline{p}',\underline{x}')\in \mathit{Tr}$ be a transition formula of the form \eqref{eq:trans-form}.  We write $\{\varphi\}~\tau~\{\psi\}$ (in analogy with Hoare triples) to abbreviate the following formula:
\begin{eqnarray}
  \label{eq:vc-val}
  \forall \underline{x},\underline{x}'.\,
  \varphi(\underline{p},\underline{x}) \wedge
  \tau(\underline{p},\underline{x},\underline{p}',\underline{x}') 
  \rightarrow
  \psi(\underline{p}',\underline{x}') ,
\end{eqnarray}
whose validity modulo $T_\mathit{SOA}$ implies that \emph{the transition $\tau$ leads 
$\mathcal{A}$ from a state satisfying $\varphi$ to one satisfying $\psi$}.
\begin{definition}
  Let $\mathcal{A}=(\underline{x},\underline{p},\iota,Tr)$ be a two-level SO
  transition system with background theory $T_\mathit{SOA}$, let $\tau_1, ..., \tau_n$
  be a sequence of transition formulae such that $\tau_i\in Tr$, and let
  $\varphi_0, ..., \varphi_n$ be a sequence of quantifier-free state
  formulae. The \emph{(symbolic) execution problem} consists of
  checking whether $\tau_i$ leads $\mathcal{A}$ from a state
  satisfying $\varphi_i$ to a state satisfying $\varphi_{i+1}$, or,
  equivalently, to checking
  \begin{eqnarray*}
    T_\mathit{SOA} & \models & \{\varphi_i\}~\tau~\{\varphi_{i+1}\} 
  \end{eqnarray*}
  for each $i=0, ..., n-1$. \hfill \close
\end{definition}

\begin{property}
  \label{prop:pre-comp-se}
  Let $\varphi$ and $\psi$ be two quantifier-free state formulae and
  $\tau$ be a transition.  Then, it is possible to effectively compute
  a quantifier-free formula $\phi$ that is logically equivalent to the
  negation of $\{ \varphi \} ~\tau~ \{ \psi\}$ and such that
  $T_\mathit{SOA}\models \{ \varphi \} ~\tau~ \{ \psi\}$ iff $\phi
  \mbox{ is $T_\mathit{SOA}$-unsatisfiable.}$ \hfill \close
\end{property}

\begin{SHORT}
  That is, for quantifier-free $\varphi$ and $\psi$, the negation of
  $\{ \varphi \} ~\tau~ \{ \psi\}$ ``is'' still quantifier-free.
\end{SHORT}
\begin{LONG}
  That is, for quantifier-free $\varphi$ and $\psi$, the negation of
  $\{ \varphi \} ~\tau~ \{ \psi\}$ ``is'' still quantifier-free (e.g.,
  the negation of $\{ \iota \}~ \mathit{GetRoleCertEmpl} ~ \{
  \varphi_1\}$ in the case study in Appendix~\ref{subsec:running-ex}).
\end{LONG}

If we are able to check the $T_\mathit{SOA}$-satisfiability of
quantifier-free formulae, then we are also able to solve the symbolic
execution problem for the two-level SO transition system with
$T_\mathit{SOA}$ as background theory. We now identify sufficient
conditions on the component theories of $T_\mathit{SOA}$ (i.e.\
$T_\mathit{sub}, T_\mathit{WF},$ and $T_\mathit{PM}$) for the
decidability of the $T_\mathit{SOA}$-satisfiability problem.
\begin{lemma}
  \label{th:dec-SOA-sat}
  Let $T_\mathit{sub}$ be an enumerated data-type theory, and $T_\mathit{WF} \supseteq
  T_\mathit{sub}$ and $T_\mathit{PM} \supseteq T_\mathit{sub}$ be consistent theories with
  decidable satisfiability problems.  The $T_\mathit{SOA}$-satisfiability
  problem is decidable for $T_\mathit{SOA}= T_\mathit{WF}\cup T_\mathit{PM}$.
  \hfill \close
\end{lemma}

\begin{SHORT}
Pragmatical aspects of how to implement the decision procedures are discussed in~\cite{BRV-TR09}.
\end{SHORT}
\begin{LONG}
Pragmatical aspects of how to implement the decision procedures are discussed in Appendix~\ref{sec:pragmatics-executability}, extending the observations on the pragmatics of modeling WF and PM of SO applications of Appendix~\ref{sec:pragmatics-modeling}.
\end{LONG}

We are now in the position to state the main result of this section, which follows from the properties and lemmas above.
\begin{theorem}
  Let $T_\mathit{sub}$ be an enumerated data-type theory, and $T_\mathit{WF} \supseteq
  T_\mathit{sub}$ and $T_\mathit{PM} \supseteq T_\mathit{sub}$ be consistent theories with
  decidable satisfiability problems.  Then, the symbolic execution
  problem for two-level SO transition systems with background theory
  $T_\mathit{SOA}=T_\mathit{WF}\cup T_\mathit{PM}$ is decidable. \hfill \close
\end{theorem}

Note that the use of an enumerated data-type theory as $T_\mathit{sub}$
does not imply that only two-level SO transition systems with finite state space can
be verified by our method.  In fact, both $T_\mathit{WF}$ and $T_\mathit{PM}$ can
have models with infinite cardinalities (this is the case, for
example, of the theory $\mathit{MsgPass}$).  So, symbolic execution is
decidable even if the state space of the two-level SO transition systems is infinite
provided that there exist decision procedures for the theories
characterizing the WF and the PM levels and it is possible to
find a common sub-theory, used for synchronization by the two levels,
whose models are finite.  

\subsection{Invariant verification of SO applications}
\label{sec:inv-so-apps}

Recall that we fixed a two-level SO transition system
$\mathcal{A}=(\underline{x},\underline{p},\iota,Tr)$ with background
theory $T_\mathit{SOA}$.  We now consider the problem of verifying that 
$\mathcal{A}$ satisfies a certain security property
$\phi$, in symbols $\mathcal{A} \models \phi$.  Since many interesting
security properties can be expressed as invariance properties (e.g., for the verification of security protocols or web services), which are a sub-class of safety properties, we assume
below that $\phi$ is a state formula of the form
\begin{eqnarray}
  \label{eq:inv-form}
  \forall \underline{i}:\mathit{Id}.\, \varphi(\underline{i},\underline{x},\underline{p}) .
\end{eqnarray}
Two remarks are in order.  First, we are considering a sub-class of
invariance properties since, in general, $\varphi$ can be a
past-formula (see, e.g.,~\cite{manna-pnueli-book}).  Second, we cannot
assume that a finite and known number of principals is fixed so that
(\ref{eq:inv-form}) is equivalent to a quantifier-free formula and thus
the verification techniques in~\secref{sec:exec-so-apps} still apply.
Rather, we want to verify that for a fixed but unknown number of
principals, $\mathcal{A}$ satisfies the invariance
property $\phi$, i.e.\ we want to solve a \emph{parameterized}
invariance verification problem.  For this reason, in the rest of this
section, we assume that:
\begin{verification-assumption}
$T_\mathit{sub}$ is the theory of an equivalence relation. \hfill \close
\end{verification-assumption}

In this way, we are able to distinguish between the identifiers of the various principals.  
Now, in order to show that formulae of the form (\ref{eq:inv-form}) are
invariant of $\mathcal{A}$, we can use the well-known INV rule of
Manna and Pnueli~\cite{manna-pnueli-book}:
\begin{eqnarray*}
  \begin{array}{lrcl}
    (I_1) &  T_\mathit{SOA} & \models &
            \forall \underline{i}.\iota(\underline{i})
                                   \rightarrow \psi(\underline{i}) \\
    (I_2) & T_\mathit{SOA} & \models &
            \forall \underline{i}.\psi(\underline{i})
                                   \rightarrow \varphi(\underline{i}) \\
    (I_3) & T_\mathit{SOA} & \models & \{ \psi\}~ \tau~ \{ \psi\} \mbox{ for each } \tau\in \mathit{Tr}\\
    \hline \\[-1em]
          & \mathcal{A} & \models & \Box \varphi
  \end{array}
\end{eqnarray*}
The intuition underlying the correctness of the rule is the following.
Assume there exists a formula $\psi$ of the form
(\ref{eq:inv-form}) identifying a set of states that includes both
the set of initial states $(I_1)$ and the set of states characterized
by $\varphi$ $(I_2)$, and, furthermore is an invariant of
$\mathcal{A}$ $(I_2)$, i.e.\ each transition of $\tau_i$ in
$\mathit{Tr}$ leads from a state satisfying $\psi$ to a state
satisfying again $\psi$.  Then, also $\varphi$ is an invariant of $\mathcal{A}$.  

Using the INV rule, assuming that the invariant $\psi$ has been
guessed, it is possible to reduce the problem of verifying that a
certain property is an invariant of the application, to several
$T_\mathit{SOA}$-satisfiability problems.  In fact, reasoning by
contradiction we have that $(I_1)$ and $(I_2)$ hold iff the
quantifier-free formulae
\begin{eqnarray*}
  \iota(\underline{i},\underline{p},\underline{x}) \wedge 
  \neg \psi(\underline{i},\underline{p},\underline{x}) \quad \mbox{ and } \quad
  \psi(\underline{i},\underline{p},\underline{x})  \wedge 
  \neg \varphi(\underline{i},\underline{p},\underline{x})
\end{eqnarray*}
are $T_\mathit{SOA}$-unsatisfiable, respectively, where the variables in
$\underline{i}$ are regarded as (Skolem) constants.  Similarly,
for a given $\tau$, $(I_3)$ holds iff the (universally) quantified
formula
\begin{eqnarray*}
  \forall \underline{j}.\, \psi(\underline{j},\underline{p},\underline{x}) \wedge
  \tau(\underline{p},\underline{x},\underline{p}',\underline{x}') 
  \wedge
  \neg \psi(\underline{i}, \underline{p}',\underline{x}') 
\end{eqnarray*}
is $T_\mathit{SOA}$-unsatisfiable, where $\underline{x},\underline{x}'$ and
$\underline{i}$ are regarded again as (Skolem) constants, for each
$\tau\in \mathit{Tr}$.  
\begin{property}\label{property:sat}
  Let $\psi_0(\underline{x},\underline{p}) :=\forall
  \underline{i}.\,\psi(\underline{i},\underline{x},\underline{p})$ be a
  state formula and
  $\tau(\underline{x},\underline{p},\underline{x}',\underline{p}')$ be
  a transition formula.  Then, it is possible to effectively compute a
  formula of the form
  $\psi_1'(\underline{i},\underline{x},\underline{p})\wedge \forall
  \underline{j}.\,\psi_2'(\underline{i},\underline{x},\underline{p})$
  that is logically equivalent to the negation of $\{ \psi_0 \}
  ~\tau~ \{ \psi_0 \}$ and such that
   $T_\mathit{SOA}\models \{ \psi_0 \} ~\tau~ \{ \psi_0 \}$ iff 
    $\psi_1'(\underline{i},\underline{x},\underline{p})\wedge 
    \forall \underline{j}.\psi_2'(\underline{i},\underline{x},\underline{p})
    \mbox{ is $T_\mathit{SOA}$-unsatisfiable.}$
\hfill \close
\end{property}

To be able to check that $\mathcal{A} \models \Box \varphi$, we need
to solve the $T_\mathit{SOA}$-satisfiability problem of (universally)
quantified formulae.  We now identify sufficient conditions on the
background theory $T_\mathit{SOA}$ for this problem to be decidable.
\begin{lemma}
  \label{th:dec-SOA-sat-univ}
  Let $\Sigma_\mathit{sub}$ contain only (countably many) constant symbols
  (i.e.\ $T_\mathit{sub}$ is the theory of an equivalence relation), $T_\mathit{WF}
  \supseteq T_\mathit{sub}$ be a consistent and stably-infinite theory with
  decidable satisfiability problem such that the signature of no
  function symbol in $\Sigma_\mathit{WF}$ is of the kind $S_1 \times \cdots
  S_n \rightarrow \mathit{Id}$ (for $S_i \in \Sigma_\mathit{WF}$ and
  $i=1,...,n$) and $T_\mathit{PM} \supseteq T_\mathit{sub}$ be a consistent BSR
  theory.  Then, the $T_\mathit{SOA}$-satisfiability problem is decidable for
  formulae of the form
  \begin{eqnarray*}
    \forall \underline{i}:\mathit{Id}.\,
     \varphi(\underline{i},\underline{x},\underline{p}) ,
  \end{eqnarray*}
  where $T_\mathit{SOA}= T_\mathit{WF}\cup T_\mathit{PM}$ and $\underline{x},\underline{p}$
  are (finite) sequences of variables and predicate symbols (such that
  $\underline{p} \cap \Sigma_\mathit{SOA}=\emptyset$), respectively.
  \hfill \close
\end{lemma}

As we have already said, there are many theories formalizing data
structures (such as $\mathit{MsgPass}$) relevant
for modeling the WF of SO applications, which are stably
infinite.  Also, it is frequently the case that the data structures
formalized by these theories use identifiers to create new pieces of
information (typical examples are the certificates of the identities or
the roles of principals) but do not create new identifiers.  In this
way, the requirement that functions in $T_\mathit{WF}$ do not create
identifiers (syntactically, this is expressed by forbidding that the
return type of the functions is not $\mathit{Id}$) is frequently
satisfied.  We point out that this requirement is a sufficient
condition to avoid the creation of new identifiers and it may be
weakened.  However, we leave the study of more general sufficient
conditions to future work. 
\begin{SHORT}
Pragmatical aspects of invariant verification of SO applications are
discussed in~\cite{BRV-TR09}.
\end{SHORT}
\begin{LONG}
As before, pragmatical aspects of invariant verification of SO applications are
discussed in Appendix~\ref{sec:pragmatics-invariant}, extending
appendices~\ref{sec:pragmatics-modeling} and~\ref{sec:pragmatics-executability}.
\end{LONG}

We conclude this section with the main technical result, which follows from
the above properties and lemma.
\begin{theorem}
  Let $\Sigma_\mathit{sub}$ contain only (countably many) constant symbols
  (i.e.\ $T_\mathit{sub}$ is the theory of an equivalence relation), $T_\mathit{WF}
  \supseteq T_\mathit{sub}$ be a consistent and stably-infinite theory with
  decidable satisfiability problem such that the signature of no
  function symbol in $\Sigma_\mathit{WF}$ is of the kind $S_1 \times \cdots
  S_n \rightarrow \mathit{Id}$ (for $S_i \in \Sigma_\mathit{WF}$ and
  $i=1,...,n$), and $T_\mathit{PM} \supseteq T_\mathit{sub}$ be a consistent BSR
  theory.  Let
  $\mathcal{A}=(\underline{x},\underline{p},\iota,\mathit{Tr})$ be a
  two-level SO transition system with background theory 
  $T_\mathit{SOA}=T_\mathit{WF} \cup T_\mathit{PM}$, and
  $\forall \underline{i}.\, \varphi(\underline{i},\underline{x},\underline{p})$ be
  a state formula.  It is decidable to check whether $\mathcal{A}
  \models \forall
  \underline{i}.\, \varphi(\underline{i},\underline{x},\underline{p})$,
  provided there exists a state formula $\forall
  \underline{i}.\, \psi(\underline{i},\underline{x},\underline{p})$ such
  that (a) $T_\mathit{SOA} \models \forall \underline{i}.\iota(\underline{i})
  \rightarrow \psi(\underline{i})$, (b) $T_\mathit{SOA} \models \forall
  \underline{i}.\psi(\underline{i}) \rightarrow
  \varphi(\underline{i})$, and (c) $T_\mathit{SOA} \models \{ \psi\}~ \tau~
  \{ \psi\} \mbox{ for each } \tau\in \mathit{Tr}$. \hfill \close
\end{theorem}

Indeed, the usefulness of the theorem depends on the availability of
the formula $\forall \underline{i}.\, \psi(\underline{i})$, which is
called an \emph{inductive} invariant since it is preserved under the
application of the transitions of the two-level SO transition system.  Since the
problem of finding such a formula when $\underline{p}=\emptyset$ is
undecidable (see, e.g.,~\cite{manna-pnueli-book}), it is also
undecidable in our case.  However, several heuristics have been
proposed; see, e.g.,~\cite{bradley-manna}
for a recent proposal and pointers to the literature.  An interesting
line of future work is to adapt these techniques to find invariants of
SO applications.  Note that it is possible to dispense with the
computation of the auxiliary invariant whenever $\forall
\underline{i}.\, \varphi(\underline{i})$ is already inductive; in which
case, conditions (a) and (b) of the theorem are trivially satisfied and
we are only required to discharge proof obligation (c).

\section{Related work and conclusions}\label{sec:conclusions}

We have presented a two-level formal framework that allows us to specify and verify the interplay of authorization policies and workflow in SO applications and architectures. 
In the previous sections, we already discussed relevant related works
and also pointed out different research lines along which we are
currently extending the techniques and results presented here.  In
particular, as we remarked, formal methods are being increasingly
applied extensively to support the correct design of SO
applications. These works range from extending the workflow with
access control aspects (e.g.,
\cite{BertinoCramptonPaci06,PaciBertinoCrampton08}) to, vice versa,
embedding the workflow within the access control system (e.g.,
\cite{SecPAL-homepage, fisler06,DKAL-homepage,ryan05}), thus mainly
focusing on one level at a time and abstracting away most or all of
the possible interplay between the WF and PM levels. Other works
(e.g. \cite{armando-ponta09,bcm-aimsa08,BalbianiChevalierElHouri-framework,schaad})
have in contrast proposed approaches that attempt to model and analyze
the interplay. We believe that our framework is abstract enough to
encompass such approaches and we are currently investigating how they
can be recast in our framework. In particular, we plan to use our
framework to model Petri nets and access control policies as
in~\cite{armando-ponta09} so as to perform deductive-based model
checking of security-sensitive business processes, and also to
formally analyze properties of RBAC by adapting the framework
of~\cite{bcm-aimsa08,BalbianiChevalierElHouri-framework}.  Finally, we
also plan to extend the framework as we presented it in this paper,
e.g.~with interfaces more refined than the $T_\mathit{sub}$ we
considered here, so 
as to be able to perform modular reasoning in the assume-guarantee
style.

\subsubsection*{Acknowledgments}
The work presented in this paper
was partially supported by the FP7-ICT-2007-1 Project no.~216471,
``AVANTSSAR: Automated Validation of Trust and Security of
Service-oriented Architectures" and the PRIN'07 project ``SOFT''. We
thank the members of the AVANTSSAR project for useful discussions.



\begin{LONG}
\newpage

\appendices

\section{Proofs}

\begin{IEEEproof}[Proof of Property~\ref{prop:pre-comp-se}]
  We reason by contradiction and reduce validity to satisfiability.
  The validity (modulo $T_\mathit{SOA}$) of formulae of the form
  (\ref{eq:vc-val}) is equivalent to the $T_\mathit{SOA}$-unsatisfiability of
  the negation of (\ref{eq:vc-val}), i.e.
  \begin{eqnarray*}
    \exists \underline{x},\underline{x}'.\,
    \varphi(\underline{p},\underline{x}) \wedge
    \tau(\underline{p},\underline{x},\underline{p}',\underline{x}') 
    \wedge
    \neg \psi(\underline{p}',\underline{x}') ,
  \end{eqnarray*}
  which, in turn, is equivalent to the $T_\mathit{SOA}$-unsatisfiability of
  \begin{multline*}
    \exists \underline{x},\underline{x}'.\,
    \varphi(\underline{p},\underline{x}) \wedge
    \exists \underline{i},\underline{d}.\ (
    G(\underline{i}, \underline{d}) \,\wedge\,
    \underline{x}' = \underline{f}(\underline{x}, \underline{i}) \,\wedge\\
    \qquad \forall \underline{z}.\, \underline{p}'(\underline{z}) \leftrightarrow
    \underline{\phi}(\underline{z}, \underline{i}) )
    \wedge
    \neg \psi(\underline{p}',\underline{x}') .
  \end{multline*}
  After some simple logical manipulations, the problem reduces to
  checking the $T_\mathit{SOA}$-unsatisfiability of the following
  quantifier-free formula
  \begin{multline*}
    \varphi(\underline{p},\underline{x}) \wedge
    G(\underline{i}, \underline{d}) \,\wedge\,
    \underline{x}' = \underline{f}(\underline{x}, \underline{i}) \,\wedge\\
    \qquad
    \forall \underline{z}.\, \underline{p}'(\underline{z}) \leftrightarrow
    \underline{\phi}(\underline{p}, \underline{z}, \underline{i}) 
    \wedge
    \neg \psi(\underline{p}',\underline{x}') ,
  \end{multline*}
  where $\underline{x},\underline{x}'$ are considered as Skolem
  constants (or, equivalently, as implicitly existentially quantified
  variables).  Now, to simplify our argument and make it easier to
  grasp, we use a little bit of higher-order logic and regard $\forall
  \underline{z}.\, \underline{p}'(\underline{z}) \leftrightarrow
  \underline{\phi}(\underline{p},\underline{z},\underline{i})$ as
  $\underline{p}' = \lambda
  \underline{z}.\, \underline{\phi}(\underline{p},\underline{z},\underline{i})$.
  In this way, it is obvious that, after two simple substitutions, the
  last formula above becomes
  \begin{eqnarray*}
    \varphi(\underline{p},\underline{x}) \wedge
    G(\underline{i}, \underline{d}) \,\wedge\,
    \neg \psi(\lambda \underline{z}.\,\underline{\phi}(\underline{p},\underline{z},\underline{i}),\underline{f}(\underline{x}, \underline{i})) ,
  \end{eqnarray*}
  which is easily seen to be equisatisfiable to the previous one
  (intuitively, it is always possible to find an assignment for the
  variables in $\underline{x}'$ when the last formula is satisfiable:
  just take the values of $\underline{f}(\underline{x},
  \underline{i})$; a similar observation holds for the predicate
  symbols in $\underline{p}'$).  Now, we are left with the problem of
  checking whether the last formula is quantifier-free.  To see this,
  recall that $\psi$ is quantifier-free and this implies that every
  occurrence of a predicate symbol in $\underline{p}'$ is applied to a
  tuple of ground terms.  Hence, the effect of substituting
  $\underline{p}'$ with $\lambda
  \underline{z}.\, \underline{\phi}(\underline{p},\underline{z},\underline{i})$
  is a tuple of quantifier-free formulae because of $\beta$-reduction.
  Let
  $\overline{\psi}(\underline{\phi}(\underline{p},\underline{i}),\underline{f}(\underline{x},
  \underline{i}))$) be the result of exhaustively performing such
  $\beta$-reductions; then, the formula
  \begin{eqnarray*}
    \label{eq:qf-vc}
    \varphi(\underline{p},\underline{x}) \wedge
    G(\underline{i}, \underline{d}) \,\wedge\,
    \neg \overline{\psi}(\underline{\phi}(\underline{p},\underline{i}),\underline{f}(\underline{x}, \underline{i})) . 
  \end{eqnarray*}
  is quantifier-free.  This concludes the proof. 
\end{IEEEproof}

\begin{IEEEproof}[Proof of Lemma~\ref{th:dec-SOA-sat}]
  We apply one of the results on non-disjoint combination of theories
  in~\cite{ghilardi-jar}, namely the following: if (i) $T_\mathit{sub}$ is a
  universal theory contained in both $T_\mathit{WF}$ and
  $T_\mathit{PM}$, (ii) $T_\mathit{sub}$ admits a model completion
  $T_\mathit{sub}^*$, (iii) every model of $T_\mathit{WF}$ and
  $T_\mathit{PM}$ embeds into a model of $T_\mathit{WF}\cup T_\mathit{sub}^*$
  and of $T_\mathit{WF}\cup T_\mathit{sub}^*$, respectively, and (iv)
  $T_\mathit{sub}$ is effectively locally finite, then the
  $(T_\mathit{WF}\cup T_\mathit{PM})$-satisfiability problem is
  decidable (by an extension of the Nelson-Oppen combination schema).
  Let us check each of the conditions (i)--(iv):
  \begin{itemize}
    \item[(i)] This is satisfied by assumption.
    \item[(ii)] By a well-known result in model-theory (see,
      e.g.,~\cite{chang-kiesler}), a theory $T$ admitting elimination
      of quantifiers also admits a model completion $T^*$ and
      furthermore $T=T^*$.  It is not difficult to check that
      $T_\mathit{sub}$ admits elimination of quantifiers (it is sufficient to
      note that $\exists x.\, \varphi(x)$ is $T_\mathit{sub}$-equivalent to
      $\bigvee_{c_i} \varphi(x/c_i)$ for $\varphi$ a quantifier-free
      formula and $c_i$'s the constants denoting the elements of the
      domain of the enumerated data type).  Hence, $T_\mathit{sub}$ admits a
      model completion and $T_\mathit{sub}^*=T_\mathit{sub}$.
    \item[(iii)] Since $T_\mathit{sub}^*=T_\mathit{sub}$ and, by assumption,
      $T_\mathit{WF}\supseteq T_\mathit{sub}$ and $T_\mathit{PM}\supseteq T_\mathit{sub}$, we have
      that $T_\mathit{WF}\cup T_\mathit{sub}^*=T_\mathit{WF}\cup T_\mathit{sub}=T_\mathit{WF}$ and,
      similarly, $T_\mathit{PM}\cup T_\mathit{sub}^*=T_\mathit{PM}\cup T_\mathit{sub}=T_\mathit{PM}$.
      This implies that every model of $T_\mathit{WF}$ (respectively, $T_\mathit{PM}$) is also a
      model of $T_\mathit{WF}\cup T_\mathit{sub}^*$ (respectively, $T_\mathit{WF}\cup T_\mathit{sub}^*$),
      which implies that every model of $T_\mathit{WF}$
      (respectively, $T_\mathit{PM}$) can be embedded into a model of $T_\mathit{WF}\cup
      T_\mathit{sub}^*$ (respectively, $T_\mathit{WF}\cup T_\mathit{sub}^*$), just take
      identity as the embedding.\footnote{An embedding $\mu$ between
        two $\Sigma$-structures $\mathcal{M}=(M,{I})$ and
        $\mathcal{N}=(N,{J})$ is a mapping from $M$ to $N$ such that
        $\mathcal{M} \models \alpha$ iff $\mathcal{N} \models \alpha$,
        for every $\Sigma^{M}$-atom $\alpha$.  (In other words, an
        embedding between $\mathcal{M}$ and $\mathcal{N}$ is an
        isomorphism of $\mathcal{M}$ onto a sub-structure of
        $\mathcal{N}$.)  We say that $\mathcal{M}$ is \emph{embeddable} in
        $\mathcal{N}$ if there exists an embedding between
        $\mathcal{M}$ and $\mathcal{N}$.}
    \item[(iv)] Indeed, the signature of an enumerated data-type
      theory is finite and consists of a finite set of constants, say
      $\{c_1, ..., c_n\}$ for some $n\geq 1$.  The constants $c_1,
      ..., c_n$ are the representatives since, for every term $t$,
      $T_\mathit{sub}\models t=c_i$ for some $i$.\footnote{A $\Sigma$-theory
        $T$ is \emph{locally finite} if $\Sigma$ is finite and, for every set
        of constants $\underline{a}$, there are finitely many ground
        terms $t_1, ..., t_{k_{\underline{a}}}$, called
        \emph{representatives}, such that for every ground
        $\Sigma^{\underline{a}}$ -term $u$, we have $T\models u = t_i$
        for some $i$.  If the representatives are effectively
        computable from $\underline{a}$ and $t_i$ is computable from
        $u$, then $T$ is effectively locally finite.}   
  \end{itemize}
\end{IEEEproof}

\begin{IEEEproof}[Proof of Property~\ref{property:sat}]
  Reason by refutation and expand the definition of the negation of
  $\{ \psi_0\} \tau \{ \psi_0\}$, so as to obtain the following
  formula:
  \begin{eqnarray*}
    \exists \underline{x},\underline{x}'.\,
    \forall \underline{i}.\, \psi(\underline{i},\underline{p},\underline{x}) \wedge
    \tau(\underline{p},\underline{x},\underline{p}',\underline{x}') 
    \wedge
    \neg \forall \underline{i}.\, \psi(\underline{i}, \underline{p}',\underline{x}') .
  \end{eqnarray*}
  This, in turn, is equivalent to
  \begin{eqnarray*}
    \exists \underline{x},\underline{x}', \underline{j}.\,
    \forall \underline{i}.\, \psi(\underline{i},\underline{p},\underline{x}) \wedge
    \tau(\underline{p},\underline{x},\underline{p}',\underline{x}') 
    \wedge
    \neg \psi(\underline{j}, \underline{p}',\underline{x}').
  \end{eqnarray*}
  Then, by recalling the definition of $\tau$ and performing the
  obvious substitutions (in a way similar to what we have done in the
  proof of Property~\ref{prop:pre-comp-se} above), we obtain:
  \begin{eqnarray*}
    \exists \underline{x},\underline{x}', \underline{j}, \underline{k}, \underline{d}.\left(
      \begin{array}{l}
        \forall \underline{i}.\, \psi(\underline{i},\underline{p},\underline{x}) \wedge 
        G(\underline{k}, \underline{d}) \,\wedge\, \\
        \underline{x}' = \underline{f}(\underline{x}, \underline{k})
        \,\wedge\,
        \underline{p}' = \lambda
        \underline{z}.\, \underline{\phi}(\underline{p},\underline{z},\underline{k}) \wedge \\
        \tau(\underline{p},\underline{x},\lambda
        \underline{z}.\, \underline{\phi}(\underline{p},\underline{z},\underline{k}),\underline{f}(\underline{x}, \underline{k})) \wedge \\
        \neg \psi(\underline{j}, 
        \lambda \underline{z}.\,
        \underline{\phi}(\underline{p},\underline{z},\underline{k}),
        \underline{f}(\underline{x}, \underline{k}))
      \end{array} \right) ,
  \end{eqnarray*}
  which, by exhaustively applying $\beta$-reduction and considering
  existentially quantified variables as Skolem constants, is
  equivalent to
  \begin{multline*}
    \forall \underline{i}.\psi(\underline{i},\underline{p},\underline{x})
    \wedge 
    G(\underline{k}, \underline{d}) \,\wedge\, \\
    \overline{\tau}(\underline{p},\underline{x},
    \underline{\phi}(\underline{p},\underline{z},\underline{k}),
    \underline{f}(\underline{x}, \underline{k})) \wedge 
    \neg \overline{\psi}(\underline{j}, 
    \underline{\phi}(\underline{p},\underline{z},\underline{k}),
    \underline{f}(\underline{x}, \underline{k})) ,
  \end{multline*}
  where $\overline{\tau}$ and $\overline{\psi}$ are the result of
  $\beta$-reducing the corresponding formulae.  It is not difficult to
  see that the last formula is a conjunction of a universally
  quantified formula $\forall
  \underline{i}.\, \psi(\underline{i},\underline{p},\underline{x})$ with
  a quantifier-free formula $
  \overline{\tau}(\underline{p},\underline{x},
  \underline{\phi}(\underline{p},\underline{z},\underline{k}),
  \underline{f}(\underline{x}, \underline{k})) \wedge 
  \neg \overline{\psi}(\underline{j},
  \underline{\phi}(\underline{p},\underline{z},\underline{k}),
  \underline{f}(\underline{x}, \underline{k}))$ and that it cannot be
  simplified further or, in other words, that the universal
  quantifier on $\underline{i}$ cannot be removed.  This concludes the
  proof. 
\end{IEEEproof}

\begin{IEEEproof}[Proof of Lemma~\ref{th:dec-SOA-sat-univ}]
  We claim that the quantifier-free formula 
  \begin{eqnarray*}
    \bigwedge_{\sigma} 
     \varphi(\underline{i}\sigma,\underline{x},\underline{p}) 
  \end{eqnarray*}
  is $T_\mathit{SOA}$-equisatisfiable to the universally quantified formula
  above, where $\sigma$ ranges over all possible ground substitutions
  mapping the variables in $\underline{i}$ to a finite subset of
  constant symbols in $\Sigma_\mathit{sub}$.  If the $T_\mathit{SOA}$-satisfiability
  problem is decidable (for quantifier-free) formulae, then the proof
  is complete.  To this end, we use the same combination result
  in~\cite{ghilardi-jar} as that for Lemma~\ref{th:dec-SOA-sat}, i.e.\ if
  (i) $T_\mathit{sub}$ is a universal theory contained in both $T_\mathit{WF}$ and
  $T_\mathit{PM}$, (ii) $T_\mathit{sub}$ admits a model completion $T_\mathit{sub}^*$,
  (iii) every model of $T_\mathit{WF}$ and $T_\mathit{PM}$ embeds into a model of
  $T_\mathit{WF}\cup T_\mathit{sub}^*$ and of $T_\mathit{WF}\cup T_\mathit{sub}^*$, respectively,
  and (iv) $T_\mathit{sub}$ is effectively locally finite, then the
  $(T_\mathit{WF}\cup T_\mathit{PM})$-satisfiability problem is decidable (by an
  extension of the Nelson-Oppen combination schema).  Let us check
  each of the conditions (i)--(iv):
  \begin{itemize}
  \item[(i)] $T_\mathit{sub}$ is the theory of an equivalence relation, which
    can be axiomatized by a finite set of universal sentences
    corresponding to reflexivity, symmetry, and transitivity.  Hence,
    $T_\mathit{sub}$ is universal.
  \item[(ii)] It is well-known that the model-completion $T_\mathit{sub}^*$
    of the theory of an equivalence relation is the theory of an
    infinite set (see, e.g.,~\cite{ghilardi-jar}).
  \item[(iii)] It is possible to show that this is equivalent to
    stably infiniteness (again, see~\cite{ghilardi-jar}) which is an
    assumption for $T_\mathit{WF}$ while it can be easily shown that any BSR
    theory is stably infinite (see, e.g.,~\cite{tinelli-zarba}), hence
    $T_\mathit{PM}$ is also so.
  \item[(iv)] $T_\mathit{sub}$ is the theory of an equivalence relation with
    finitely many equivalence classes.  So, although there are
    infinitely (more precisely, countably many) constant symbols of
    sort $\mathit{Id}$, there exists a finite subset $C=\{c_1, ...,
    c_n\}$ such that for any other constant symbol $d$ of sort
    $\mathit{Id}$, we have $T_\mathit{sub} \models d=c_i$ for some $i\in\{1,
    ..., n\}$.  This means that $T_\mathit{sub}$ is an effectively locally
    finite theory.
  \end{itemize}
  Thus, we conclude that the $T_\mathit{SOA}$-satisfiability problem for
  quantifier-free formulae is decidable.  

  To conclude the proof, we are left with the problem of proving the
  claim above.  To this end, first of all, recall that $T_\mathit{sub}$ is
  effectively locally finite.  Then, observe that $T_\mathit{PM}$ is a BSR
  theory and hence $\Sigma_\mathit{PM}$ has no function symbol of arity
  greater than $0$.  Furthermore, recall that, by assumption, all
  function symbols of arity greater than $0$ in $\Sigma_\mathit{WF}$ are such
  that their return type is not $\mathit{Id}$.  Thus, we have that
  $T_\mathit{PM} \models d=c_i$ and $T_\mathit{WF} \models d=c_i$, for every
  constant symbol of sort $\mathit{Id}$ in $\Sigma_\mathit{sub}$ and some
  $c_i\in C$.  This is so because the reduct to $\Sigma_\mathit{sub}$ of
  every $\Sigma_\mathit{PM}$-model of $T_\mathit{PM}$ and $\Sigma_\mathit{WF}$-model of
  $T_\mathit{WF}$ is a model of $T_\mathit{sub}$.  So, if the quantifier-free formula
  \begin{eqnarray*}
    \bigwedge_{\sigma} 
     \varphi(\underline{i}\sigma,\underline{x},\underline{p}) 
  \end{eqnarray*}
  is $T_\mathit{SOA}$-unsatisfiable, where $\sigma$ is a ground substitution
  mapping the variables in $\underline{i}$ to the computable finite
  subset $C$ of the constants in $\Sigma_\mathit{sub}$ of sort
  $\mathit{Id}$, then the universally quantified formula
  \begin{eqnarray*}
    \forall \underline{i}:\mathit{Id}.\,
     \varphi(\underline{i},\underline{x},\underline{p}) ,
  \end{eqnarray*}
  is also unsatisfiable.  For the converse, it is sufficient to recall
  that $T_\mathit{sub}$ is a universal theory and that universal theories are
  closed under sub-structures, i.e.\ any sub-structure of a model of
  the theory is also a model.  This implies that if the
  quantifier-free formula above is $T_\mathit{SOA}$-satisfiable, then the
  universally quantified-formula is also so. 
\end{IEEEproof}

\section{A case study: Car Registration Office}
\label{subsec:running-ex}

The techniques and results that we give in this paper are general and independent of particular concrete applications, but, to illustrate them concretely, it is useful to consider an example from industrial practice: we consider a simplified version of the car registration office case study described in~\cite{avantssar-deliverable-5.1}, which can be intuitively summarized as follows.

A \emph{citizen}, called Charlie, submits a request to register his new
car to an \emph{employee}, called Ed, of the local \emph{car registration office CRO}.\footnote{We have abstracted away the mechanism assigning a citizen request to a certain employee of the car
 registration scenario.} Charlie's message contains all the documents to
support his request and it is suitably signed.  Upon reception of the
request, Ed has appropriate support for checking the signature of
the document and comparing it with the identity of the sender of the
request: if the signature and the identity of the requester do not
match, then the request is immediately refused and the sender is
acknowledged of this fact; otherwise, Ed starts to consider the
content of the request for the car registration. If, according to some
criteria (that are abstracted away in the specification), the request
is not suitably supported by the documents, then the request is
refused and, again, the sender is acknowledged of this fact; otherwise, the request is 
accepted, the sender is acknowledged of
acceptance and the request is marked as accepted, signed by Ed, and
finally sent to the \emph{central repository CRep} to be archived.

This process is completely transparent to Charlie and, in order to be
successfully completed, Ed should have the right to store documents
in the \emph{CRep}.  This right 
can only be granted by the \emph{head} of the \emph{CRO}, 
a (special) employee called Helen.
Upon reception of the request by Ed to store a processed request in
its internal database, the \emph{CRep} checks whether Ed has
been granted the right to do so. If this is the case, the \emph{CRep}
stores the document; otherwise, it refuses to comply.

Roles are assigned to employees (of the \emph{CRO}) by
circulating appropriate certificates; such as, e.g., ``Ed is an
employee'' or ``Helen is the head of the \emph{CRO}.''
These certificates are emitted by a \emph{certification authority RegOffCA}, that 
is recognized by the employees of the \emph{CRO} and the \emph{CRep}. Permission to store documents in the
\emph{CRep} are also distributed to employees by creating
appropriate certificates; however, these certificates are created by
the head of the \emph{CRO} (not by the certification authority).  

The \emph{CRep}, before storing a processed request in its
internal database, checks whether the employee has the right to do
so. For this to be successfully executed, the following policy should
be enforced:
\begin{itemize}
\item an employee of the car registration office can store documents
  in the \emph{CRep}, if the head of the car registration
  office permits it,
\end{itemize}
and the following trust relationships should have been preliminarily
established:
\begin{itemize}
\item the \emph{RegOffCA} 
is trusted by all employees, by the head of the car registration
  office, and the \emph{CRep} for what concerns role certificates; and
\item the head of the car registration office is trusted by the
  \emph{CRep} for action (e.g., storing documents) certificates.
\end{itemize}
Finally, to be able to successfully execute the scenario with Charlie and
Ed described above, the following certificates should be available
in the system:
\begin{itemize}
\item Ed is an employee of the car registration office (by a
  certificate emitted by \emph{RegOffCA}),
\item Helen is the head of the car registration office (by a
  certificate emitted by \emph{RegOffCA}), and 
\item Helen permits Ed to store documents in the \emph{CRep} (by a certificate emitted by Helen).
\end{itemize}

\subsection*{Formalization}
Since only the exchange of messages drives the workflow of the system
and the most interesting part of the case study concerns its policies, 
we adopt the $\mathit{MsgPass}[\mathit{Msg}]$ theory described in Example \ref{ex:msg-pass}. 
In the body of a message, documents (such as car
registration requests or processed requests) can be embedded
($\mathsf{embeddoc}$).  Since both citizens and employees should be able
to sign documents and the latter should also be able to check
signatures, appropriate primitives  are provided to generate signatures
($\mathsf{sign}$), attaching them to documents
($\mathsf{augdocwithsign}$), and checking that the signature attached to
a document belongs to a certain principal ($\mathsf{matchuser}$).  An employee has also the primitive to attach a decision
($\mathtt{accept}$ or $\mathtt{refuse}$) to a document containing a
citizen request ($\mathsf{augdocwithact}$).  Finally, as role
certificates should be distributed over the network, we provide an
appropriate primitive ($\mathsf{rolecert}$) to create these documents.
(Role certificates are handled at the policy level only; see below for
more details.)
Now, we identify the theories involved:
\begin{itemize}
\item $T_\mathit{sub} = \\
\mathit{EDT}(\{\mathtt{Charlie}, \mathtt{Ed} , \mathtt{Helen}, \mathtt{CRep}, \mathtt{RegOffCA}\}, \mathit{Id}) ~ \cup \\
\mathit{EDT}(\{\mathtt{employee}, \mathtt{head} \} , \mathit{Role}) ~ \cup \\
\mathit{EDT}(\{\mathtt{storedoc}, \mathtt{readdoc} \} , \mathit{Action})$
\item $T_\mathit{WF} = T_\mathit{sub} \cup \mathit{MsgPass}[\mathit{Msg}]$, the theory described in the Example \ref{ex:msg-pass} above,
\item $T_\mathit{PM} = T_\mathit{sub} ~\cup$ \\
  $\{Knowledge_{0\infty}, Say2know_{0\infty},
  Trustedknowledge_{0\infty}\}$, a set of Horn rules defined
  below.\footnote{The class of Horn rules is an extension of that of
    Datalog rules whereby function symbols of arity greater than $0$
    are allowed.}
\end{itemize}
As we have said above, the workflow of the system is almost state-less.
There are however two exceptions.  One is the database of the central
repository which is modeled by the unary predicate $\mathsf{dbdoc}$ to
which documents may only be added (and never deleted).  The other is
the unary predicate $\mathsf{isok}$ that allows us to abstract away the
criteria according to which a citizen request is accepted or refused.
This completes the description of the (static part of the) workflow.

We now describe the policy level of the system.  We adapted
the DKAL~\cite{DKAL-homepage} approach to specifying policies in our framework.  To
this end, DKAL provides predicates ($\mathsf{knows}$ and
$\mathsf{knows}_0$) to represent the knowledge of the various agents and
predicates ($\mathsf{saysTo}$ and $\mathsf{saysTo}_0$) for the
\emph{communication} between agents.\

It is important to observe the differences between the communication
at the workflow and the policy levels of the system. The former
(modeled via the $\mathit{MsgPass}[\mathit{Msg}]$ theory) is state-full
and thus modeled by an appropriate set of transitions (see below).
The latter (modeled via $\mathsf{saysTo}$ or $\mathsf{saysTo}_0$) is
state-less and thus modeled by suitable Horn clauses.  Finally, DKAL
proposes two functions ($\mathsf{tdOn}$ or $\mathsf{tdOn}_0$) to track
trust relationships between agents concerning certain facts.  All this
is formally captured in the following set of Horn clauses.

First, we provide an (incomplete) characterization of the DKAL-like predicates expressing
knowledge and communication for policies (this is adapted
from~\cite{DKAL-homepage}, to which the interested reader is pointed to for
details).  
\\

\noindent
$Knowledge_{0\infty}$: Internal knowledge is knowledge.
\begin{small}\begin{align*}
\mathsf{knows}(P,AnyThing) \leftarrow \mathsf{knows}_{0}(P,AnyThing) 
\end{align*}\end{small}

\noindent
$Say2know_{0\infty}$: An agent knows whatever is said to him and he/she also knows whether the piece 
of knowledge being communicated is based on the internal knowledge of the 
speaker ($say2know_0$) or not ($say2know_{\infty}$).

\begin{small}\begin{align*}
\mathsf{knows}_{0}(P,said0(Q,AnyThing)) \leftarrow \mathsf{saysTo}_{0}(Q,AnyThing,P)\\
\mathsf{knows}(P,said(Q,AnyThing)) \leftarrow \mathsf{saysTo}(Q,AnyThing,P)
\end{align*}\end{small}

\noindent
$Trustedknowledge_{0\infty}$: An agent $P$ knows a piece of information $AnyThing$ whenever 
an agent $P$ knows that another agent $Q$ said the piece of 
information $AnyThing$ and also that $P$ knows that the agent $Q$
is trusted on saying the piece of information $AnyThing$.\

\begin{small}\begin{align*}
\mathsf{knows}(P,AnyThing) \leftarrow & \mathsf{knows}(P,\mathsf{tdOn}_{0}(Q,AnyThing)) ~ \wedge \\
											                  & \mathsf{knows}(P,\mathsf{said}_{0}(Q,AnyThing)) \\
\mathsf{knows}(P,AnyThing)  \leftarrow & \mathsf{knows}(P,\mathsf{tdOn}(Q,AnyThing)) ~ \wedge \\
                                                               & \mathsf{knows}(P,\mathsf{said}(Q,AnyThing))
 \end{align*} \end{small}

Then, we consider the policies of each agent.  The first three Horn
clauses specify the communication of (role and action) certificates at
the policy level.  (Note that while the knowledge of an action
certificate for $\mathtt{Ed}$ is explicitly given in the initial state above,
the knowledge of the role certificates for $\mathtt{Ed}$ and $\mathtt{Helen}$ will be
lifted from the existence of the corresponding messages in the network
by appropriate transitions.)  The last three Horn clauses are the
formal counterparts of the trust relationships described above.\\

\noindent
\textbf{(Simple) employee }
  
\begin{small}\begin{align}
\mathsf{saysTo}(Empl,& \mathsf{said}_{0} (\mathtt{RegOffCA},Cert),\_) \leftarrow \tag{Cert1}\\
& \mathsf{knows}(Empl,\mathsf{said}_{0}(\mathtt{RegOffCA},Cert)) \notag \\
\mathsf{saysTo}(Empl,& \mathsf{said}_{0} (Head,Cert),\_) \leftarrow \tag{Cert2} \\
& \mathsf{knows}(Empl,\mathsf{said}_{0}(Head,Cert)) \notag
\end{align}\end{small}

In the Horn rules above, and also in some of the following ones, we use a
Prolog-like notation where the symbol $\_$ is employed as an abbreviation
for a universally quantified variable that occurs only once in the rule.\\

\noindent
\textbf{(Head) employee}

\begin{small}\begin{align*}
\mathsf{saysTo}_{0} & (Head,\mathsf{storedocCRep}(Empl,\_)  \leftarrow \notag \\
& \mathsf{knows}_{0}(Head,\mathsf{storedocCRep}(Empl)) \tag{GenCert} 
\end{align*}\end{small}

\noindent
\textbf{Central Repository } \\
 
\noindent
\begin{small}$\mathsf{knows}(\mathtt{CentrRep},\mathsf{tdOn}_{0}(\mathtt{RegOffCA},\_))$ \hfill (CentrRepTrustCA)\\\end{small}

\noindent
\begin{small}$\mathsf{knows}(\mathtt{CentrRep},\mathsf{tdOn}(\_,\mathsf{said}_{0}(\mathtt{RegOffCA},\_)))$
\begin{flushright}
(CentrRepTrustAnyoneViaCA)
\end{flushright}
\end{small}

\begin{small} \begin{align*}
&\mathsf{knows} (\mathtt{CentrRep},\mathsf{tdOn}_{0}(Head,\mathsf{storedocCRep}(Empl))) \leftarrow \notag \\
& \mathsf{knows}(\mathtt{CentrRep},\mathsf{said}_{0}(\mathtt{RegOffCA},\mathsf{isRegOffHead}(Head))) ~\wedge \notag \\
& \mathsf{knows}(\mathtt{CentrRep},\mathsf{said}_{0}(\mathtt{RegOffCA},\mathsf{isRegOffEmpl}(Empl))) \\
\tag{CentrRepTrustAnyoneViaHead}
 \end{align*}\end{small}

Finally, in Fig.~\ref{fig:transition}, we give the transitions modeling the dynamics of the system.
The first two transitions ($\mathit{GetRoleCertEmpl}$, $\mathit{GetRoleCertHead}$), are part of the interface between the
workflow and the policy levels of the system as they allow employees
to convert the content of role certificates received from the network
to (internal) knowledge, which is relevant for the application of
policies (compare the right-hand sides of these rules with the
hypotheses of the Horn clauses Cert1 and Cert2).
The following two transitions specify the processing of a citizen
request by an employee ($\mathit{Accept}$), and how the central repository handles the request of an
employee to store a document in its internal database ($\mathit{Storedoc}$). This is (the
remaining) part of the interface between the workflow and the policy
level: the guard

\begin{small}     
$$ \mathsf{knows}(\mathtt{CentrRep},\mathsf{storedocCRep} (Empl)) $$
\end{small}

\noindent
is a query that is possibly solved by the Horn clauses above.

\begin{figure*}[t]
$\mathit{GetRoleCertEmpl}$:
 \begin{eqnarray*}
    \exists i_1, i_2. 
     \left(
       \begin{array}{l}
         \mathsf{mem}(\mathsf{msg}(\mathtt{RegOffCA}, \mu[\mathsf{rolecert}(i_1, \mathtt{employee})], i_2), \mathit{net})
         ~\wedge \
         \mathit{net}' = \mathit{net} ~ \wedge \\
         
         \forall p_1,p_2,r.
           \mathsf{hasrole}'(p_1,p_2,r) \leftrightarrow 
           \left(
             \begin{array}{l}
               \mathit{if}~ (p_2=i_1 \wedge p_1=i_2 \wedge r=\mathtt{employee}) \\
               \mathit{then ~~ true} \\
               \mathit{else} ~ \mathsf{hasrole}(p_1,p_2,r) 
             \end{array} 
           \right)
         \end{array}
       \right)
  \end{eqnarray*}
where, $\mu$ is a term symbol of type $\mathsf{Body}$ that contains a sub-term of interest that represent a role certificate.\\
\noindent
$\mathit{GetRoleCertHead}$:
 \begin{eqnarray*}
    \exists i_1, i_2. 
     \left(
       \begin{array}{l}
         \mathsf{mem}(\mathsf{msg}(\mathtt{RegOffCA}, \mu[\mathsf{rolecert}(i_1, \mathtt{head})], i_2), \mathit{net})
         ~\wedge \
         \mathit{net}' = \mathit{net} ~ \wedge \\
         
         \forall p_1,p_2,r.
           \mathsf{hasrole}'(p_1,p_2,r) \leftrightarrow 
           \left(
             \begin{array}{l}
               \mathit{if}~ (p_2=i_1 \wedge p_1=i_2 \wedge r=\mathtt{head}) \\
               \mathit{then ~~ true} \\
               \mathit{else} ~ \mathsf{hasrole}(p_1,p_2,r) 
             \end{array} 
           \right)
         \end{array}
       \right)
  \end{eqnarray*}
  \noindent
$\mathit{Accept}$:
 \begin{eqnarray*}
    \exists d,c,i. 
     \left(
       \begin{array}{l}
         \mathsf{mem}(\mathsf{msg}(c, \mathtt{embeddoc}(d), i), \mathit{net}) ~ \wedge  \mathtt{isok}(d) ~ \wedge
         \mathsf{matchuser}(d,c) ~ \wedge \\
         net'=\mathsf{ins}(\mathsf{msg}(i,\mu[\mathsf{augdocwithdec}(d,\mathtt{acceptdoc})],\mathtt{CentrRep}),net) ~ \wedge \\
         \forall p_1,p_2,r.
           \mathsf{hasrole}'(p_1,p_2,r) \leftrightarrow \mathsf{hasrole}(p_1,p_2,r)
         \end{array}
       \right)
  \end{eqnarray*}
    \noindent
$\mathit{Storedoc}$:
 \begin{eqnarray*}
    \exists i,d. 
     \left(
       \begin{array}{l}
         \mathsf{mem}(\mathsf{msg}(i, \mu[\mathsf{augmentdocwithact}(d,\mathtt{storedoc})] ,\mathtt{CentrRep}), \mathit{net})
          ~ \wedge \\
         \mathsf{dbdoc}'=\mathsf{dbdoc}(d) ~ \wedge \\
         \forall p_1,p_2,r.
           \mathsf{hasrole}'(p_1,p_2,r) \leftrightarrow \mathsf{hasrole}(p_1,p_2,r)
         \end{array}
       \right)
  \end{eqnarray*}
\caption{Transition formulae
\label{fig:transition}}%
\end{figure*}

\subsection*{Executability}
\label{ex:po-exec}
It is relatively easy to check that the scenario described above
involving Ed and Helen can be executed by a suitable sequence of
transitions and solving appropriate queries against the policies of the
system. For instance, let us, for 
%
the sake of brevity, only analyze the first step, which requires the application of the
transition $\mathit{GetRoleCertEmpl}$ to lead the two-level SO transition system from
the initial state to a state where the PM knowledge about the identity
of Ed has been acquired.  The initial state $\iota$ is characterized
by the following formula, 

\begin{small}\begin{align*}
\mathit{net} & = \\
& \mathsf{ins}(\mathsf{msg}(\mathtt{Charlie},\mathsf{embeddoc} (\mathsf{augdocwithsign}(req, \\
  & ~~~~~~~~~~ \mathsf{sign}(\mathtt{Charlie},req))), \mathtt{Ed}),\\
  & \mathsf{ins}(\mathsf{msg}(\mathtt{RegOffCA},\mathsf{embeddoc} (\mathsf{augdocwithsign}(\rho_E, \\
  & ~~~~~~~~~~\mathsf{sign}(\mathtt{RegOffCA},\rho_E))), \mathtt{Ed}), \\
  & \mathsf{ins}(\mathsf{msg}(\mathtt{RegOffCA}, \mathsf{embeddoc}(\mathsf{augdocwithsign}(\rho_H, \\
  & ~~~~~~~~~~\mathsf{sign}(\mathtt{RegOffCA},\rho_H))), \mathtt{Ed}),\mathsf{mty}))) ~ \wedge \\
  & \bigwedge_{p_1,p_2\in C, r\in R} \neg \mathsf{hasrole}(p_1,p_2,r)
\end{align*}\end{small}

\noindent
  saying that principals knows nothing about their respective roles
  and the net contains three messages: one is the car registration
  request of Charlie and the other two are the role certificates of Ed
  (who is an employee) and Helen (who is the head of the car
  registration office), where $\rho_E$ and $\rho_H$ abbreviate the
  terms $\mathsf{rolecert}(\mathtt{Ed},\mathtt{employee})$ and
  $\mathsf{rolecert}(\mathtt{Helen},\mathtt{head})$, respectively, $C=\{
  \mathtt{RegOffCA},$ $\mathtt{CRep}, \mathtt{Ed}, \mathtt{Charlie},
  \mathtt{Helen}\}$, and $R=\{\mathtt{employee}, \mathtt{head}\}$.

  The transition $\mathit{GetRoleCertEmpl}$ is formalized as in Fig.~\ref{fig:transition}.
  The set of states to which the transition $\mathit{GetRoleCertEmpl}$
  should lead the two-level SO transition system must be so as to satisfy the
  following formula $\varphi_1$:
  
\begin{small}  \begin{eqnarray*}
    \mathsf{knows}(\mathtt{Ed}, \mathsf{isRegOffEmpl}(\mathtt{Ed}))
  \end{eqnarray*}\end{small}

\noindent
  saying that $\mathtt{Ed}$ has acquired the knowledge about its role
  at the PM level of the SO application.  It is not difficult to show
  the $T_\mathit{SOA}$-validity of $\{\iota\}~ \mathit{GetRoleCertEmpl}~
  \{\varphi_1\}$.  In fact, the transition is enabled since the
  following formula (obtained by instantiating both $i_1$ and $i_2$
  with the constant $\mathtt{Ed}$ and substituting the state variable
  $\mathit{net}$ with the term at the right of the first equality 
  in $\iota$):
  
 \begin{small} \begin{align*}
  \mathsf{mem}(\mathsf{msg} & (\mathtt{RegOffCA}, \mathsf{embeddoc} \\
  & (\mathsf{augdocwithsign}(\rho_E, \mathsf{sign}(\mathtt{RegOffCA},\rho_E))),\mathtt{Ed})), \\  
  & ~~~\mathsf{ins}(\mathsf{msg}(\mathtt{Charlie},\mathsf{embeddoc} (\mathsf{augdocwithsign}(req, \\
  & ~~~~~~~~~~ \mathsf{sign}(\mathtt{Charlie},req))), \mathtt{Ed}),\\
  & ~~~\mathsf{ins}(\mathsf{msg}(\mathtt{RegOffCA},\mathsf{embeddoc} (\mathsf{augdocwithsign}(\rho_E, \\
  & ~~~~~~~~~~\mathsf{sign}(\mathtt{RegOffCA},\rho_E))), \mathtt{Ed}), \\
  & ~~~\mathsf{ins}(\mathsf{msg}(\mathtt{RegOffCA}, \mathsf{embeddoc}(\mathsf{augdocwithsign}(\rho_H, \\
  & ~~~~~~~~~~\mathsf{sign}(\mathtt{RegOffCA},\rho_H))), \mathtt{Ed}), \\
  & \mathsf{mty}))))
  \end{align*}\end{small}

\noindent
  is $T_\mathit{SOA}$-satisfiable.  We are left with the problem of showing
  the $T_\mathit{SOA}$-unsatisfiability of the formula:
  \begin{small}\begin{equation*}
    \neg \mathsf{knows}(\mathtt{Ed}, \mathsf{isRegOffEmpl}(\mathtt{Ed}))
  \end{equation*}\end{small}

\noindent
  obtained by negating $\varphi_1$.  This can be easily done by
  observing that 

  \begin{small}\begin{equation*}
    \mathsf{hasrole}(\mathtt{Ed}, \mathtt{Ed}, \mathtt{employee})
  \end{equation*}\end{small}

\noindent
  holds in the state where the transition $\mathit{GetRoleCertEmpl}$ has
  lead the two-level SO transition system as the result of executing the PM update.
  Then, by instantiating the following Horn rule (in $T_\mathit{PM}$):

\begin{small}  \begin{eqnarray*}
    \mathsf{knows}(i_1,\mathsf{isRegOffEmpl}(i_2)) 
    & \leftarrow &
    \mathsf{hasrole}(i_1, i_2, \mathtt{employee})
  \end{eqnarray*}\end{small}

\noindent
  with $i_1$ and $i_2$ substituted with $\mathtt{Ed}$, it is possible
  to immediately detect unsatisfiability.  In this way, we have proved
  the $T_\mathit{SOA}$-unsatisfiability of the formula $\neg (\{ \iota \}~
  \mathit{GetRoleCertEmpl} ~ \{ \varphi_1\})$ or, equivalently, the
  $T_\mathit{SOA}$-validity of $\{ \iota \}~ \mathit{GetRoleCertEmpl} ~ \{
  \varphi_1\}$. \hfill \close

\subsection*{Invariant properties}
We consider the following interesting property about documents stored
in the central repository:
\begin{description}
\item[Integrity:] ~~ any processed request $\mathit{preq}$ stored in the
  central repository must be consistent, i.e., it should be double
  signed (by the citizen $\mathit{cit}$ submitting the request
  $\mathit{req}$ and by the employee $\mathit{empl}$ handling it) and
  stamped with the seal of acceptance.
\end{description}
Such a property can be written as the following safety formula in
the extended version of LTL introduced above:

{\small
\begin{eqnarray*}
  \begin{array}{l}
  \Box \left(
  \begin{array}{l}
    \forall \mathit{preq}. \mathtt{dbdoc}(\mathit{preq})
    \Rightarrow 
    \exists \mathit{cit}, \mathit{req}, \mathit{empl}, 
            \mathit{preq}_1, \mathit{preq}_2. \\
     \left(
      \begin{array}{lcll}
        {\mathit{preq}_1}  =  
        \mathsf{augdocwithsign}(\mathit{req}, 
                                  \mathtt{sign}(\mathit{user}, \mathit{req})) ~ \wedge \\
        {\mathit{preq}_2}  = 
        \mathtt{augdocwithdec}({\mathit{preq}_1},
           \mathtt{accept})  ~ \wedge \\
        \mathit{preq} =
        \mathsf{augdocwithsign}({\mathit{preq}_2},
          \mathtt{sign}(\mathit{empl},{\mathit{preq}_2})) 
       \end{array}                                 
     \right)  
    \end{array}
    \right)
    \end{array}
   \end{eqnarray*}
 } 

 Showing that the SO application ensures integrity is non-trivial, as
 the central repository treats documents as black-boxes and trusts
 employees to check signatures and correctly prepare processed
 requests.  Furthermore, it trusts the head of the central repository
 to judge the capability of employees to perform this job correctly.
 Ultimately, the central repository also trusts the certification
 authority to emit role certificates for both employees and the head
 of the car registration office.  Besides these difficulties, the state
 formula inside the ``always-in-the-future'' operator is not of the
 kind supported by the decidability result of
 Lemma~\ref{th:dec-SOA-sat-univ} because of the existential
 quantifier.  As a consequence, more ingenuity is required by the
 specifier.  We are currently working to derive a hand proof of this
 property.

\section{Pragmatics}

\subsection{Pragmatics of modeling WF and PM of SO applications}\label{sec:pragmatics-modeling}
We extend the technical results of Section~\ref{sec:modelling} with some observations on the pragmatics of modeling WF and PM of SO applications. In fact, pragmatically, the theories $T_\mathit{WF}$ and $T_\mathit{PM}$ are obtained by
extending the substrate theory $T_\mathit{sub}$ as follows.  For the WF
theory, consider a finite set $Ax(WF)$ of universal
$\Sigma_\mathit{WF}$-sentences where $\Sigma_\mathit{WF}\supseteq \Sigma_\mathit{sub}$;
then
\begin{eqnarray*}
  T_\mathit{WF} & := & \{ \psi ~\mbox{is a sentence} ~|~ 
                   Ax(WF)\models \psi \} \cup T_\mathit{sub} .
\end{eqnarray*}
The process of adding finitely many axioms to an available theory can
be iterated several times to obtain the final WF theory.  As an
example, recall the theories $\mathit{Msg}$ and 
$\mathit{MsgPass}[\mathit{Msg}]$ of Examples~\ref{ex:msg-pass} and~\ref{ex:SO-appl}.
For the PM theory, along the lines of several other works in the PM
literature (e.g.,~\cite{SPKI-SDSI-in-FOL}), regard a logic program
$P(PM)$ (formalizing policy statements) as a set of universal Horn
$\Sigma_\mathit{PM}$-clauses where $\Sigma_\mathit{PM} := \Sigma_\mathit{sub} \cup
\underline{R}$ for $\underline{R}$ a (finite) set of predicate symbols
such that $\Sigma_\mathit{sub}\cap \underline{R}=\emptyset$;\footnote{For the
  sake of conciseness, $\Sigma_\mathit{sub}\cup \underline{R}$ will be
  usually abbreviated with $\Sigma_\mathit{sub}^{\underline{R}}$ with the implicit 
  assumption that $\underline{R}$ is disjoint from
  $\Sigma_\mathit{sub}$.} then
\begin{eqnarray*}
  T_\mathit{PM} & := & \{ \psi ~\mbox{is a Horn clause} ~|~ 
                   P(PM)\models \psi \} \cup T_\mathit{sub} .
\end{eqnarray*}
Usually, the state predicates in $P(\mathit{PM})$ 
are intensional, i.e.\ occur in the
head of the rules of $P(PM)$.  This is a sufficient condition to
ensure that no transition may add a fact to the theory $T_\mathit{PM}$ that
gives rise to an inconsistency.

In the (constraint) logic programming literature, $T_\mathit{sub}$ is usually
introduced as a certain first-order structure (e.g., the integers).
This is not compatible with the notion of theory adopted here (and in
most logic textbooks) as we work with sets of sentences (axioms)
rather than structures.  However, given a structure $\mathcal{M}$,
it is possible to find a theory $T$ admitting $\mathcal{M}$ as a
model.  So, if we are able to verify that $T\models \varphi$, we also
know that $\mathcal{M}\models \varphi$ (while the converse, in
general, may not hold).  As a consequence, if we are able to reduce a
certain verification problem for an SO application to showing that a
formula follows from the background theory of the application and we
succeed in doing this, we are entitled to conclude that the
verification problem has a positive answer for any structure
satisfying the axioms of the background theory.  Indeed, we may obtain
false negatives, as there may exist formulae that are true in a
particular model of a theory $T$ that are not logical consequences of
$T$.  An advantage of adopting this notion of theory is the
possibility of re-using and adapting existing automated reasoning
techniques (see below).

\subsection{Pragmatics of executability of SO applications}\label{sec:pragmatics-executability}

Recall our remark on how the theories $T_\mathit{WF}$ and
$T_\mathit{PM}$ are formed by augmenting the theory $T_\mathit{sub}$ with a finite
set of universal axioms (see end of~\secref{sec:modelling}), i.e.\
\begin{eqnarray*}
    T_\mathit{WF} & := & \{ \psi ~\mbox{is a sentence} ~|~ 
                   Ax(WF)\models \psi \} \cup T_\mathit{sub}  \mbox{ and }\\
   T_\mathit{PM} & := & \{ \psi ~\mbox{is a Horn clause} ~|~ 
                   P(PM)\models \psi \} \cup T_\mathit{sub} ,
\end{eqnarray*}
where $Ax(WF)$ is a (finite) set of universal sentences and $P(PM)$ is
a (finite) set of Horn clauses.  It is not difficult to argue that
both the $T_\mathit{WF}$- and $T_\mathit{PM}$-satisfiability problems are decidable.
The decidability of the former can be derived by the decidability of
the $\mathit{MsgPass}$-satisfiability problem (shown in~\cite{ic03}), the
decidability of the satisfiability problem of any enumerated data-type
theory (since it admits elimination of quantifiers), and the
combination results in~\cite{cade08}.  It is possible to use available
SMT solvers (such as Yices, Z3, or MathSAT) to obtain a decision
procedure for the $T_\mathit{WF}$-satisfiability problem (almost)
off-the-shelf; maybe using characteristic functions for sets and then
using arrays of Booleans to formally represent such functions, see,
e.g.,~\cite{bradley-manna}.  The decidability of the
$T_\mathit{PM}$-satisfiability is an immediate consequence of the
(well-known) decidability of the satisfiability problem for BSR
theories.  Since $T_\mathit{sub}$ is an enumerated data-type theory, the
hypotheses of Lemma~\ref{th:dec-SOA-sat} are satisfied and we are
entitled to conclude the decidability of the $T_\mathit{SOA}$-satisfiability
problem.  Again, it is possible to use available SMT solvers, such as
Z3, to have direct support for the class of BSR theories and hence to
implement a decision procedure for the $T_\mathit{PM}$-satisfiability
problem.

We are then left with the problem of modularly reusing the decision
procedures for the satisfiability problem in the component theories to
obtain a decision procedure for the $T_\mathit{SOA}$-satisfiability problem.
When $T_\mathit{sub}$ is an enumerated data-type theory, as it is the case of
Lemma~\ref{th:dec-SOA-sat}, it is possible to use the
non-deterministic version of the combination algorithm
in~\cite{ghilardi-jar} to implement a decision procedure for the
$T_\mathit{SOA}$-satisfiability problem.  To understand this, let us briefly
summarize an adaptation of the non-deterministic combination schema
of~\cite{ghilardi-jar}.  To this end, let w.l.o.g.\ $\Gamma$ be a
conjunction of $\Sigma_\mathit{SOA}$-literals.\footnote{Given a
  quantifier-free $\Sigma_\mathit{SOA}$-formula, it is always possible to
  transform this into disjunctive normal form, i.e.\ into a
  disjunction of conjunctions of literals.  Hence, being able to check
  the satisfiability of conjunctions of literals is sufficient to
  check the satisfiability of quantifier-free formulae.  Although,
  this is not efficient (as the transformation to disjunctive normal
  form may yield an exponentially large formula), it is sufficient
  theoretically.}  First of all, we transform $\Gamma$ into an
equisatisfiable conjunction $\Gamma_\mathit{WF}\wedge \Gamma_\mathit{PM}$ by naming
sub-terms by means of additional constants $\underline{a}$: this
process is usually called \emph{purification} and it can be implemented
in polynomial time.  As there are only finitely many constants $c_1,
..., c_n$ in the enumerated data-type theory $T_\mathit{sub}$, we
non-deterministically guess an \emph{arrangement} $\Delta$, i.e.\ a
conjunction of literals such that, for each $a\in \underline{a}$,
either $a=c_i$ or $a\neq c_i$, for each $c_i$ is in $\Sigma_\mathit{sub}$.
Then, we check whether both $\Gamma_\mathit{WF}\cup \Delta$ is
$T_\mathit{WF}$-satisfiable and $\Gamma_\mathit{PM}\cup \Delta$ is
$T_\mathit{PM}$-satisfiable.  If, for some arrangement, both tests are
successful, then we conclude the $T_\mathit{SOA}$-satisfiability of
$\Gamma_\mathit{PM}\cup \Gamma_\mathit{WF}$ (and hence of $\Gamma$); otherwise, if,
for all arrangements, the tests are negative, then we are entitled to
conclude the $T_\mathit{SOA}$-unsatisfiability of $\Gamma_\mathit{PM}\cup
\Gamma_\mathit{WF}$ (and hence of $\Gamma$).  Since the number of arrangements
is finite (one can only generate finitely many distinct equalities or
disequalities between two finite sets of constant symbols, namely
$\underline{a}$ and $c_1, ..., c_n$), the method terminates and thus
yields a decision procedure for $T_\mathit{SOA}$.

There are two problems with the combination algorithm sketched above.
First, it requires to transform quantifier-free formulae into
disjunctive normal form.  This is unacceptable for many practical
problems.  Second, the algorithm is non-deterministic and we must
refine it to obtain an implementation.  To circumvent both of these
problems, we sketch in Fig.~\ref{fig:smt-based-dec-proc-SOA} an
algorithm that can be easily implemented on top of (most) SMT solvers
and is inspired by the delayed theory combination method
of~\cite{delayed-ic}.
\begin{figure}[tb]
  \begin{center}
  \begin{minipage}{.45\textwidth}
      \begin{tabbing}
        foo \= foo \= \kill
        \textbf{function} $T_\mathit{SOA}\mbox{\texttt{-sat}}(\varphi ~:~ \mbox{quantifier-free}~\Sigma_\mathit{SOA}\mbox{-formula})$ \\
        1 \> $(\phi, \underline{a}) \longleftarrow \mathtt{purify}(\varphi)$  \\
        2 \> $A \longleftarrow \mathtt{Atoms}(\phi)\cup \mathtt{IE}(\underline{a},\underline{c})$ \\
        3\> \textbf{while} \texttt{Bool-sat}$(\phi)$ \textbf{do}\\
        4\>\> $\Gamma_\mathit{WF} \wedge \Gamma_\mathit{PM} \wedge \Delta_\mathit{sub} \longleftarrow \mathtt{pick\_total\_assign}(A, \phi)$ \\
        5\>\> $(\rho_\mathit{WF}, \pi_\mathit{WF}) \longleftarrow T_\mathit{WF}\mbox{\texttt{-sat}}(\Gamma_\mathit{WF} \wedge \Delta_\mathit{sub})$ \\
        6\>\> $(\rho_\mathit{PM}, \pi_\mathit{WF}) \longleftarrow T_\mathit{PM}\mbox{\texttt{-sat}}(\Gamma_\mathit{PM} \wedge \Delta_\mathit{sub})$ \\
        7\>\> \textbf{if} $(\rho_\mathit{WF} = \mathtt{sat} \wedge \rho_\mathit{PM} = \mathtt{sat})$ \textbf{then return} $\mathtt{sat}$ \\
        8\>\> \textbf{if} $\rho_\mathit{WF} = \mathtt{unsat}$ \textbf{then} $\phi \longleftarrow \phi \wedge \neg \pi_\mathit{WF}$ \\
        9\>\> \textbf{if} $\rho_\mathit{PM} = \mathtt{unsat}$ \textbf{then} $\phi \longleftarrow \phi \wedge \neg \pi_\mathit{PM}$ \\
        10\> \textbf{end while} \\
        11\> \textbf{return} $\mathtt{unsat}$ \\
        12 \textbf{end function}
      \end{tabbing}
    \end{minipage}
  \end{center}
  \caption{\label{fig:smt-based-dec-proc-SOA}An SMT-based decision procedure for $T_\mathit{SOA}$-satisfiability}
\end{figure}
The algorithm in Fig.~\ref{fig:smt-based-dec-proc-SOA} is an
abstraction of the so-called lazy SMT solvers.  Before entering the
main loop, the input quantifier-free formula $\varphi$ is purified
into the formula $\phi$; the function $\mathtt{purify}$ also returns
the set $\underline{a}$ of constants used for purification.  Then, the
set $A$ of atoms is formed: it is the union of the atoms occurring in
the purified formula $\phi$ and all possible equalities between the
constants in $\underline{a}$ and the constants in $\underline{c}$
(coming the underlying enumerated data-type theory $T_\mathit{sub}$) as
computed by the function $\mathtt{IE}$.  The idea underlying the main
loop of the algorithm is the following.  A \emph{theory solver} for
$T$ is any procedure capable of establishing whether any given finite
conjunction of $\Sigma$-literals is $T$-satisfiable or not.  The
\emph{lazy approach} to build SMT solvers consists of integrating a
DPLL Boolean enumerator with a theory solver (see,
e.g.,~\cite{sebastiani} for details).  Given a quantifier-free formula
$\phi$, one checks if it satisfiable by considering its atoms as
Boolean variables (cf.\ \texttt{Bool-sat} at line 3).  If it is not
the case, then we exit the main loop and return unsatisfiability of
the input quantifier-free formula (cf.\ line 11).  Otherwise, we enter
the main loop and we consider a satisfying Boolean assignment, i.e.\ a
set of literals that makes $\phi$ true when atoms are considered as
Boolean variables (cf.\ $\mathtt{pick\_total\_assign}$, line 4).
Note that a Boolean assignment consists not only of the atoms in
$\phi$ (cf.\ $\mathtt{Atoms}$ at line 2) but also of all possible
equalities between the constants in $\underline{a}$ and the constants
in $\underline{c}$ (cf.\ $\mathtt{IE}$ at line 2).  In this way, we
are guaranteed to consider all possible arrangements as defined by the
non-deterministic algorithm sketched above.  Then, we
check---separately---the $T_i$-satisfiability of the conjunction of
$\Sigma_i$-literals $\Gamma_i\wedge \Delta_\mathit{sub}$ (cf.\ lines 5 and
6): the $T_i\mbox{\texttt{-sat}}$ procedure besides returning
$\mathtt{sat}$ or $\mathtt{unsat}$ also returns a conjunction $\pi_i$
(called the conflict set) of $\Sigma_i$-literals, all of which also
occur in $\Gamma_i\wedge \Delta_\mathit{sub}$, which is $T_i$-unsatisfiable
(for $i\in \{\mathit\mathit{WF}, \mathit\mathit{PM}\}$).  If both satisfiability
checks are positive, then we return the satisfiability of the input
quantifier-free formula (cf.\ line 7).  Otherwise, i.e.\ if at least
one of the satisfiability checks returned \texttt{unsat}, the negation
of $\pi_i$, called a \emph{conflict clause}, is added to $\phi$ (cf.\
line 8 or 9) so as to reduce the number of Boolean assignments that
are to be considered in the main loop.  This is one of the key
ingredients (among many others, see, e.g.,~\cite{sebastiani}, for more
details) of the success of current state-of-the-art SMT solvers and it
avoids the burden of transforming quantifier-free formulae to
disjunctive normal form, although the problem indeed is NP-hard.

The correctness of the algorithm in
Fig.~\ref{fig:smt-based-dec-proc-SOA} is an immediate corollary of
Lemma~\ref{th:dec-SOA-sat} above.
\begin{property}
  Let $T_\mathit{sub}$ be an enumerated data-type theory, and $T_\mathit{WF} \supseteq
  T_\mathit{sub}$ and $T_\mathit{PM} \supseteq T_\mathit{sub}$ be consistent theories with
  $T_\mathit{WF}$\texttt{-sat} and $T_\mathit{PM}$\texttt{-sat} as decision
  procedures for their corresponding satisfiability problems.  Then,
  the function $T_\mathit{SOA}$\texttt{-sat} (depicted in
  Fig.~\ref{fig:smt-based-dec-proc-SOA}) is a decision procedure for
  the $T_\mathit{SOA}$-satisfiability problem.
\end{property}

\subsection{Pragmatics of invariant verification of SO applications}\label{sec:pragmatics-invariant}

Here the basis to implement
an algorithm for the $T_\mathit{SOA}$-satisfiability check of quantifier-free
formulae is almost the same as the function depicted in
Fig.~\ref{fig:smt-based-dec-proc-SOA}.  The main difference is in
the definition of arrangement.  In fact, we say that $\Delta_\mathit{sub}$ is
an arrangement for the theory $T_\mathit{sub}$ of an equivalence relation over
the set $\underline{a}$ of finite constants of sort $\mathit{Id}$ if,
for every pair $(d,d')$ for $d,d'\in \underline{a}$, either $c=c'\in
\Delta_\mathit{sub}$ or $c\neq c'\in \Delta_\mathit{sub}$.  To implement this
definition of arrangement, it is sufficient to replace line $2$ of the
function in Fig.~\ref{fig:smt-based-dec-proc-SOA} with the following
one:
\begin{eqnarray*}
  2' ~~~~ A \longleftarrow \mathtt{Atoms}(\phi) \cup \mathtt{IE}(\underline{a},\underline{a}) .
\end{eqnarray*}
Let $T_\mathit{SOA}$-\texttt{qfsat} be the new function so obtained; its
correctness is a corollary of Lemma~\ref{th:dec-SOA-sat-univ} above.
Furthermore, following the proof of Lemma~\ref{th:dec-SOA-sat-univ},
it is sufficient to generate finitely many instances of a universally
quantified formula of the form 
\begin{eqnarray*}
  \forall \underline{i}:\mathit{Id}.\,
   \varphi(\underline{i},\underline{x},\underline{p}) 
\end{eqnarray*}
to obtain a decision procedure for such a class of formulae.  Indeed,
the challenge here is to efficiently integrate the function
$T_\mathit{SOA}$-\texttt{qfsat} and an instantiation strategy for the
universally quantified variables in $\underline{i}$.  This requires
some heuristics to filter out instances that are unlikely to
contribute to detecting the unsatisfiability of the formula.  To
understand why heuristics are needed, consider that the number of the
possible ground substitutions $\sigma$ is $n^k$ where $n$ is the
length of $\underline{a}$ and $k$ is the length of $\underline{i}$.
Another key ingredient to scale up is to invoke
$T_\mathit{SOA}$-\texttt{qfsat} incrementally so as to add one by one the
instances of $\varphi$.  Since tuning these heuristics and making them
work smoothly together require extensive experimental evaluation, we
leave the details for future work.  

\end{LONG}

\end{document}